\documentclass{aastex631}
\usepackage{chemformula}
\usepackage[T1]{fontenc}
\usepackage{times, graphicx}
\usepackage{url, mhchem}

\begin{document}

\title{Exoplanet Volatile Carbon Content as a Natural Pathway for Haze Formation}

\author{Edwin A. Bergin}
\affiliation{Dept. of Astronomy, University of Michigan, Ann Arbor, MI 48104}
\author[0000-0002-1337-9051]{Eliza M.-R. Kempton}
\affiliation{Dept. of Astronomy, University of Maryland, College Park, MD 20742}
\author{Marc Hirschmann}
\affiliation{Dept. of Earth and Environmental Sciences, University of Minnesota, Minneapolis, MN 55455}
\author{Sandra T. Bastelberger}
\altaffiliation{Center for Research and Exploration in Space Science and Technology, NASA/GSFC, Greenbelt, MD 20771}
\altaffiliation{NASA GSFC Sellers Exoplanet Environments Collaboration}
\affiliation{Dept. of Astronomy, University of Maryland, College Park, MD 20742}
\author{D. J. Teal}
\affiliation{Dept. of Astronomy, University of Maryland, College Park, MD 20742}
\author{Geoffrey A. Blake}
\affiliation{Division of Geological \& Planetary Sciences, California Institute of Technology, Pasadena, CA 91125}
\author{Fred J. Ciesla}
\affiliation{Dept. of the Geophysical Sciences, The University of Chicago, Chicago, IL 60637}
\author{Jie Li}
\affiliation{Dept. of Earth and Environmental Sciences, University of Michigan, Ann Arbor, MI 48104}



\begin{abstract}

We explore terrestrial planet formation with a focus on the supply of solid-state organics as the main source of volatile carbon.  For the water-poor Earth, the water ice line, or ice sublimation front, within the planet-forming disk has long been a key focal point.  We posit that the soot line, the location where solid-state organics are irreversibly destroyed, is also a key location within the disk.  The soot line is closer to the host star than the water snowline and overlaps with the location of the majority of detected exoplanets.    In this work, we explore the ultimate atmospheric composition of a body that receives a major portion of its materials from the zone between the soot line and water ice line.   We model a silicate-rich world with 0.1\% and 1\% carbon by mass with variable water content.  We show that as a result of geochemical equilibrium, the mantle of these planets would be rich in reduced carbon but have relatively low water (hydrogen) content. 
Outgassing would naturally yield the ingredients for haze production when exposed to stellar UV photons in the upper atmosphere.   Obscuring atmospheric hazes appear common in the exoplanetary inventory based on the presence of often featureless transmission spectra \citep{Kreidberg14,Knutson14, Libby-Roberts20}.  Such hazes may be powered by the high volatile content of the underlying silicate-dominated mantle.   Although this type of planet has no solar system counterpart, it should be common in the galaxy with potential impact on habitability.   
\end{abstract}

%\keywords{tbd}

\section{Introduction} \label{sec:intro}

The compositions of bodies in the Solar System point to array of chemical environments that were present during the formation of planets and their building blocks.  Among these, a fundamental change in chemistry occurred in the solar nebula, the protoplanetary disk which circled our own sun, at so-called `snow lines'.  These chemical transitions mark the region outside of which the pressures and temperatures are such that a given molecular species would exist as a solid and inside of which that solid sublimates to the vapor.  An important consequence of these locations is that corresponding species would be abundant in solids that form outside a transition point, while these components would be scarce in solids located inside.

 Of particular importance to planet formation are the locations of the water and CO snow lines, as these have traditionally been considered to be the primary carriers of  oxygen and carbon in protoplanetary disks \citep{Oberg11_C_O}.  Importantly, the CO snow line is located many tens of astronomical units (AU) from the star \citep{Qi13_sci}, only where temperatures are $<$30 K, low enough for CO ice to form.  This would seemingly make many planets, including the known rocky exoplanets, most of which are found closer to their host star than the Earth, carbon-poor.

This picture assumes that all carbon is locked up in CO (or similarly volatile species such as CH$_{4}$ or CO$_{2}$).   However, recent work has shown that a significant amount of carbon in the interstellar medium, up to 60\% of cosmic carbon \citep{Mishra15}, is carried by refractory organics \citep{Bergin15, Gail17}.  These organics, hereafter called ``soot,'' are predominantly macro-molecular in nature and comprised of hydrocarbons/organic species \citep{Alexander13} and are the product of disequilibrium reactions in the ISM and/or outer protoplanetary disk.  These soots will remain solid at temperatures up to $\sim$500 K \citep{Li21}, and have the important property that when they are heated above their destruction temperature, they decompose into simpler, more volatile species. That is, their vaporization is irreversible.  This leads to the concept of the ``Soot Line'' in a protoplanetary disk, a location close to the star, outside of which refractory carbon would be available for incorporation into solid planetary materials, but inside of which it is absent  \citep{Kress10, Li21}.  A unique property of the soot line is that any carbon contained in vapor that mixes outward remains in the gas, and does not freeze-out again as expected around traditional snow lines \citep{Ros13}.

\begin{figure}[h]
    \centering
    \includegraphics[scale=0.6]{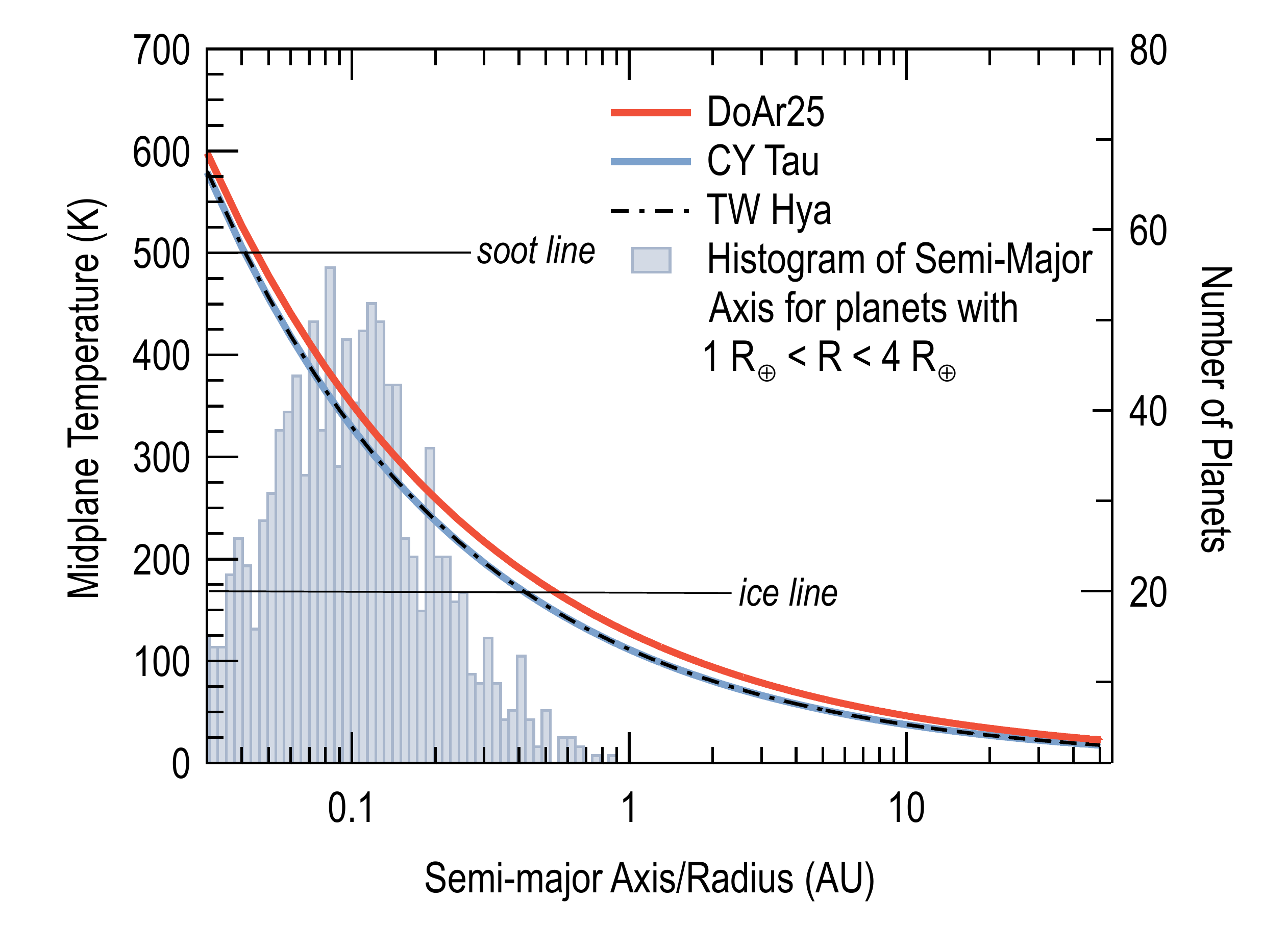}
    \caption{\bf Extrapolated midplane temperatures of mm-sized dust as a function of radius based on measurements for 3 disk systems (TW Hya and CY Tau have similar profiles and are sub-solar mass stars) \citep{Andrews16, Long18}. Also shown are current confirmed exoplanets with radius measured via primary transits.  Planets are culled to show only those with radii between 1 and 4 R$_{\oplus}$ and are plotted as function of semi-major axis and referenced to the axis on the left (Number of Planets). The soot line and ice lines are also shown which have different locations depending on the system, but the majority of detected super Earths and sub-Neptunes lie in the ``reduced carbon-rich zone''.}
    \label{fig:exo} 
\end{figure}

Planets that form outside of the soot line can thus be carbon-rich, leading to highly reducing conditions during their early evolution, particularly if they formed interior to the water snow line (and thus did not have access to another major solid hydrogen/oxygen carrier).  Fig.~\ref{fig:exo} indicates that such planets very well may exist, showing the temperature profiles of mm-sized pebbles extrapolated from ALMA measurements of protoplanetary disks of ages associated with potential incipient planet formation; the corresponding locations of the soot and snow lines are also shown.  The area in between these two locations marks the portion of the disk where planetary materials would be relatively rich in carbon but chemically reduced, because of the preservation of refractory organics but loss of water to the gas.  A histogram  of known Earth-like planets and super-Earths and their semi-major axes is also shown, with many being found in this important region.  {\em If} these planets formed predominately from solids from this region, then they would form from carbon-rich/water-poor material. It has been suggested that many of these systems formed at larger distances and migrated inwards at earlier stages \citep{Ida08, Coleman14, Izidoro17}.  This calls into question the correspondence shown in Fig.~\ref{fig:exo}.   However, other models argue for ``in situ'' formation \citep[e.g.][]{Lee14, Batygin23} and, as we discuss below, at earlier stages the nebular gas is warmer with a more distant soot line. Observational tests of these competing ideas are thus strongly desired. 

Given the above, one may expect the Earth to have formed with a high abundance of carbon, yet it is severely carbon-depleted \citep{Bergin15}.  However, there are ways that carbon can be lost from solids during planet formation.  \citet{Li21} demonstrated that temperatures early in disk evolution can be much higher than shown in Fig.~\ref{fig:exo} provided the mass accretion rate from the disk to the star is sufficiently high.  This could push the soot line out to beyond 1 au during the first 1 Myr; its low carbon inventory could reflect that much of Earth's primary materials were assembled during this phase \citep{Li21}.  Alternatively, \citet{Hirschmann21} argued that heating of planetesimals from radioactive decay ($^{26}$Al in our Solar System) can reach high enough temperatures to destroy the organics and drive off volatiles.  If Earth's building blocks formed later than suggested by \citet{Li21}, the low carbon content of the Earth could then be a result of the differentiation and thermal metamorphism of its progenitors, which we readily see in the meteorite record \citep[see also][]{Grewal2022}.

It is important to note, however, that accretion rates through disks vary by orders of magnitude \citep[e.g.][]{Hartmann16}, and thus in many protoplanetary systems the soot line would be very close to the star and well inside the location where planets are found.  Even if not, carbon-rich organic pebbles can be replenished by inward drift from regions that never saw the inside of the soot line, a process that potentially did not occur in the solar system as a result of Jupiter's formation \citep{Kruijer20}.  Further, not all planetary systems are expected to form with as high an abundance of $^{26}$Al that our Solar System had (if any at all ) \citep{Ciesla15,Lichtenberg19}, which would limit the heating that planetesimals experienced prior to their accretion into planets.  Thus, it is possible, perhaps even likely, that a significant fraction of the Earth-size planets and Super Earths were assembled from rocky materials with large carbon inventories.

 This is the thesis that we explore in this letter.  Here we focus on worlds that are silicate-rich, but have greater volatile inventories than seen in the Earth with 0.1--1.0 wt\% refractory carbon present in their mantles.   We will show that one implication of  this composition is that atmospheric hazes, which appear to be present in numerous systems \citep[e.g.][]{Kreidberg14, Crossfield17, Gao20, dymont22}, would be a natural outcome.  In \S 2 we provide the baseline model of the mantle composition.  In \S 3 we explore the atmospheric composition of these planets and apply a photochemical model of haze formation.  In \S 4 we present basic predictions of this model and discuss the implications.

  \section{Geochemical Equilibrium of Rocky Sub-Neptune Core/Terrestrial Mantle}

We  explore the consequences for the atmospheres of planets that form in this critical region where the planets are chiefly comprised of refractories and organics.   The bulk of meteoritic organic material is insoluble in typical solvents and is thought to be macromolecular in form with a typical composition of C$_{100}$H$_{75-79}$O$_{11-17}$N$_{3-4}$S$_{1-3}$ (that is, normalized relative to 100 carbon atoms) \citep{Alexander17, Glavin18}.  We assume this material is representative as the soot composition.
%It is notable that this material contains significant amounts of hydrogen.  
In the interstellar medium, up to 60\% of cosmic carbon \citep{Mishra15}, is carried by refractory organics and the bulk refractory organic carbon composition in cometary material is comparable to that of ISM material \citep{Bergin15, Gail17}.   As such, the refractory organic carbon content of planet-building materials forming beyond the soot line is expected to be high and likely comparable to that in comets.   
In this case, Comet 67P, which is similar to Comet Halley, had refractory organic carbon content that is $\sim$6$\times$ that of CI chondrites \citep{Bardyn17}, which have 2-4 wt\% carbon \citep{Pearson06}. Models of dust emission in protoplanetary disk systems also commonly assume carbonaceous dust comprised of refractory organics is present in abundances consistent with the interstellar medium \citep{Pollack94, DAlessio01,Birnstiel18}.  Further chemical evidence of soot destruction at the soot line may have been recently detected in the JWST spectra of a young protoplanetary disk \citep{Tabone23}. 

We therefore explore outcomes where the composition of the starting material is 0.1\% and 1.0\% wt\% soot, with the majority of the mass in these planets being the more refractory silicates and metals.  A final state with 0.1\% to 1\% soot by mass is conservative in that it assumes substantial volatile loss during formative stages \citep{Hirschmann21} from the assumed initial values of 12-24 wt\%. While rich in reduced carbon, these silicate-dominated rocky bodies are not the extreme ``carbon planets'' discussed by other researchers \citep[e.g.][]{Madhusudhan12, Unterborn14}. 
%Unterbore14 https://iopscience.iop.org/article/10.1088/0004-637X/793/2/124

We first assume that the planet accreted without water, as this represents an extreme case, but we also explore solutions where some water is provided in the form of hydrous silicates.  Even in the most water-poor cases, water in the mantle is generated via the oxygen in silicates and the hydrogen from organics/atmosphere; this water can be released to the atmosphere (this is discussed in Appendix).
%The ingredients for haze production have been linked to the presence of CH$_4$, and other simple hydrocarbons, that are processed by UV photons in the upper atmosphere \citep{Kempton12, Gao21}.  We note that other pathways of production have been suggested for CO and CO$_2$ dominated atmospheres\citep{He20}, alongside the potential import of trace species, e.g., sulfur \citep{Zahnle16, Gao17}. In this case we focus on CH$_4$, which we show can be quite abundant, even in cases for which temperatures would nominally convert CH$_4$ into CO or CO$_2$. 
We perform several calculations to predict the end-state compositions of nascent planetary atmospheres (see Appendix). We adopt rocky masses of 0.3~M$_\oplus$, 1.0~M$_\oplus$, and 3~M$_\oplus$.  Planets are assumed to accumulate a nebular atmosphere consisting of hydrogen that increases with their mass, following  \citet{Stokl15}, which can be significant when the planet mass exceeds that of Earth.   For all objects, we first determine the geochemical equilibrium between the outgassing from a molten mantle and the overlying atmosphere, using the outgassing model of 
\citet{Gaillard22}. Based on the assumed ratio of refractories to organics we calculate the oxygen fugacity and the compositions of coexisting atmosphere and molten volatile-bearing silicate.  In each case, the resulting mantle contains significant amounts of carbon.  The destruction of the soot results in the formation of  C-species dissolved in the molten silicate,  and in some cases, graphite, along with outgassed C-O-H vapor. 

The results from these equilibrium calculations, which represent the base of the overlying atmosphere, are given in Table~\ref{tab:geochem}.  For the lower mass planets (0.3~M$_\oplus$ and 1.0~M$_\oplus$) their initial atmospheres are dominated by H$_2$ and CO, but significant (a few \%+) amounts of CH$_4$ are typically present. The presence of a massive hydrogen envelope in the 3 M$_\oplus$ planet changes the underlying equilibrium such that the atmospheres are H$_2$ and CH$_4$ dominated.  Essentially, an appreciable fraction of C released from the mantle is processed into methane in these atmospheres. The stability of CH$_{4}$ rather than CO, even at high temperature, is a product of the high pressure of these thick atmospheres, as the reaction CO + 3 H$_{2}~{\longrightarrow}~$ CH$_{4}$ + H$_{2}$O has a negative volume change.  Surprisingly, the inclusion of water, across all modeled planetary masses, does not alter this composition as the high atmospheric carbon content continues to favor some hydrocarbon production.   These results demonstrate that abundant hydrocarbons are outgassed to the base atmosphere across a range of planet masses and even in the case of high  (compared to the Earth) water content.  This is notable as the ingredients for haze production have been linked to the presence of CH$_4$, and other simple hydrocarbons, that are processed by UV photons in the upper atmosphere \citep[e.g.][]{Kempton12, morley13, kawashima18, lavvas19}.  We note that other pathways of haze production have been suggested for CO and CO$_2$ dominated atmospheres \citep{He20}, alongside the potential importance of trace species, e.g., sulfur \citep{Zahnle16, Gao17}. In this case we concentrate on CH$_4$, which we show can be quite abundant, even in cases for which temperatures would nominally convert CH$_4$ into CO or CO$_2$.

\begin{center}
    \begin{table}
        \caption{Model Assumptions and Base Atmosphere  Properties/Composition}
        \centering
    \begin{tabular}{cccrrrcrrrrrrrrr}
    \hline
    \multicolumn{1}{c}{\% Soot$^a$} &
    \multicolumn{1}{c}{\% H$_2$O$^a$} & \multicolumn{1}{c}{M$_p$} &
 \multicolumn{1}{c}{M$_{p,soot}$} &
    \multicolumn{1}{c}{M$_{p,H_2}$} &
    \multicolumn{1}{c}{P} 
    &    \multicolumn{1}{c}{log$_{10}$(f$_{\rm O_2}$)}
    & \multicolumn{1}{c}{N$_2$} & \multicolumn{1}{c}{H$_2$O} & \multicolumn{1}{c}{H$_2$} & \multicolumn{1}{c}{CO$_2$} & \multicolumn{1}{c}{CO} & \multicolumn{1}{c}{CH$_4$} & \multicolumn{3}{c}{Element Fractions}\\
    \multicolumn{2}{c}{(by mass)} &
    \multicolumn{3}{c}{(M$_\oplus$)} & \multicolumn{1}{c}{(MPa)} 
    & \multicolumn{1}{c}{}& \multicolumn{6}{c}{(mixing ratios)} &\multicolumn{1}{c}{H}&\multicolumn{1}{c}{O}&\multicolumn{1}{c}{C}\\\hline\hline
%%soot  %h2o Planet mass (Earth Masses)	P (MPa)	N2	H2O	H2	CO2	CO	CH4
1.0 &  0.0& 0.30& 0.003 & 8.7$\times 10^{-8}$ & 84.08& -10.47 & 0.00& 0.04& 0.23& 0.03& 0.66& 0.03 &0.314	&0.351	&0.335\\
1.0 & 0.0& 1.00& 0.01 & 1.5$\times 10^{-4}$ & 98.47& -10.55 & 0.00& 0.05& 0.33& 0.02& 0.52& 0.07 &0.459	&0.271	&0.270\\
1.0 & 0.0 & 3.00& 0.03 & 1.4$\times 10^{-2}$ & 2970.00& -14.80 & 0.00& 0.00& 0.59& 0.00& 0.00& 0.41 &0.874&	0.000	&0.126\\
0.1 &  0.0 & 0.30& 0.003 & 8.7$\times 10^{-8}$ & 71.28& -10.19& 0.00& 0.00& 0.01& 0.05& 0.94& 0.00 &0.007	&0.508 &	0.484\\
0.1 &0.0 & 1.00& 0.01& 1.5$\times 10^{-4}$ & 70.52& -10.26 & 0.01& 0.01& 0.07& 0.04& 0.87& 0.00 & 0.082	&0.471	&0.446\\
0.1 & 0.0 & 3.00& 0.03 & 1.4$\times 10^{-2}$  & 866.10& -14.80 &0.00& 0.00& 0.95& 0.00& 0.00& 0.05 & 0.977	&0.000	&0.023\\
0.1 &1.0 & 0.30& 0.003 & 8.7$\times 10^{-8}$ & 109.34& -10.36 & 0.00& 0.08& 0.50& 0.01& 0.29& 0.10 & 0.668	&0.165 &	0.167\\
0.1 &1.0 &  1.00& 0.01 & 1.5$\times 10^{-4}$ & 188.16& -10.43&  0.00& 0.06& 0.42& 0.01& 0.28& 0.21 & 0.679	&0.135	&0.186\\
0.1 &1.0 &  3.00& 0.03 & 1.4$\times 10^{-2}$ &1120.00& -14.84 & 0.00& 0.00& 0.96& 0.00& 0.00& 0.04 & 0.983	&0.000	&0.017\\\hline
\multicolumn{12}{l}{$^a$Percent mass added to M$_p$.}
    \end{tabular}
    \label{tab:geochem}
\end{table}
\end{center}

\section{Atmospheric Chemical Equilibrium of Rocky Sub-Neptune Core/Terrestrial World}

\subsection{Baseline Atmospheric Composition}

\begin{figure}
    \centering
    \includegraphics[scale=0.25]{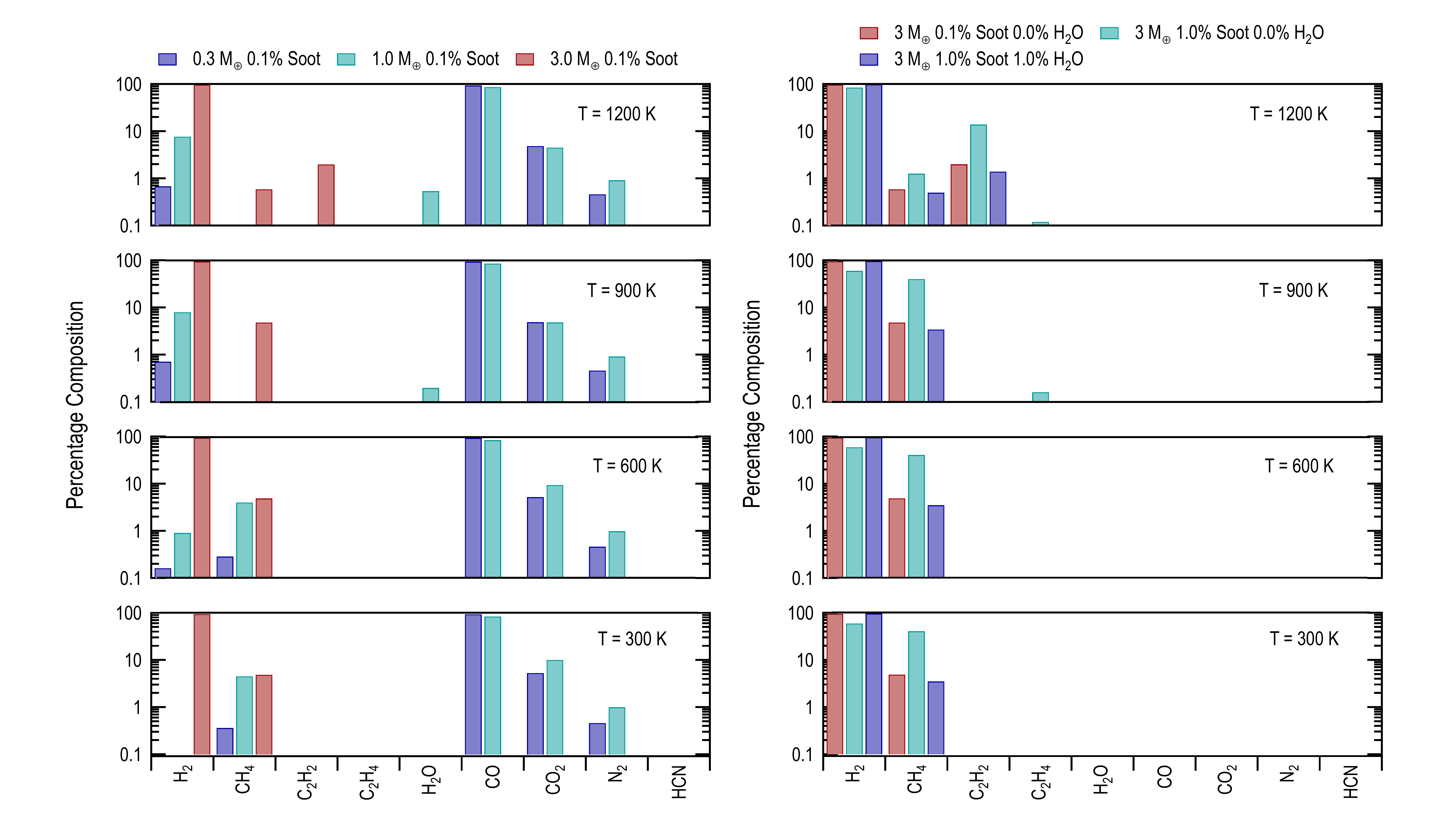}
  %%  Is left H2O free? What does 0.0 % H2O on the right mean? 
  \caption{\bf {\em Left:} Atmospheric composition for a planet with 0.1\% soot and variable mass.  These models have no additional water added beyond the water that might form via the baseline soot composition. {\em Right:} Atmospheric composition for a 3 M$_\oplus$ planet with variable soot (and water) content.  0.0\% H$_2$O refers to the baseline model where water might form from hydrogen provided by soot and 1.0\% water refers to a model with an addition 1 wt\% of water included beyond what is provided by soot.  In each case 4 temperatures are shown which effectively relates to the semi-major axis of the planet's orbit. We note that these compositions are  all at 1 mbar, which is approximately the transmission spectroscopy photosphere.  Other species (such as H$_2$S) are included in the calculation, some with abundances high enough to appear in transmission spectra. Here only those with abundances in excess of 0.1\% are shown.}
   \label{fig:atmcomp} 
\end{figure}

The results from Table~\ref{tab:geochem} refer to the composition at the base of atmosphere, but tying this to haze production requires quantitative models of chemistry and irradiation of the upper atmosphere.  To determine whether we expect high CH$_4$ abundances to persist to altitude, we first investigate the atmospheric composition in thermochemical equilibrium as a function of atmospheric temperature and pressure (methods are described in the Appendix).   The results from these calculations are provided in Fig.~\ref{fig:atmcomp}.  Here we include the atmospheric temperature as a proxy for the planet's orbital distance. The left-hand panel provides the atmospheric composition as a function of planet mass and atmospheric temperature, for planets with 0.1\% soot by mass.  For lower mass planets, a greater diversification of the carbon content is seen, resulting in CO and CO$_2$ dominated atmospheres.   Despite this, at lower temperatures some transitioning to CH$_4$ is observed, yielding significant CH$_4$/CO$_2$ ratios.  This is of interest as the simultaneous detection of CH$_4$ and CO$_2$ has been suggested as a biosignature \citep{Krissansen-Totton18}; for these planets it may instead be a natural outcome of formation.  However, for planets with $\le$1 M$_\oplus$ detailed calculations of mantle evolution are needed to understand the impact of higher soot content in the upper mantle (compared to the Earth) and the resulting effect on atmospheric composition and evolution.  

Our solutions in Table~\ref{tab:geochem} refer to young planets, but there can be significant mantle and atmospheric composition evolution over billions of years.   Most directly this would be the loss of the primary H$_2$ atmosphere and development of a secondary atmosphere.
Thus, these models are not necessarily predictive for the composition of observed, and evolved, exoplanets.  Among our modeled planets, the 3 M$_\oplus$ is dominated by its hydrogen envelope and is expected to experience the least atmospheric evolution.  We therefore focus on these planets as representative of an evolved outcome of sub-Neptune atmosphere formation. These planets are more abundant in our collection of known exoplanets and are more representative of those atmospheres likely to be characterized by JWST in the next few years. The right-hand panel of Fig.~\ref{fig:atmcomp} shows the equilibrium atmospheric composition for a 3 M$_\oplus$ planet with variable soot and water content at a pressure of 1 mbar (approximately the pressure probed by transmission spectroscopy measurements).  Here, we find that the atmospheres are hydrogen/methane dominated, and in some instances with significant concentrations of acetylene (C$_2$H$_2$) and ethylene (C$_2$H$_4$).  These results show that methane and other hydrocarbons can persist at high abundance, even in high temperature and low pressure conditions, due to the effectively elevated C/O ratio provided by the soot-rich mantle.

\subsection{Implementation of Haze Model}

Models of haze formation have been developed for exoplanetary atmospheres based upon irradiation of methane and other carriers.  We apply one such model including chemical kinetics, photochemistry, and haze formation to our 3 M$_\oplus$ planet with 0.1\% soot and no water.   We stress that this hydrocarbon-based haze model is for illustrative purposes. Our calculations show that these atmospheres will be methane rich.  But nitrogen and sulfur are carried alongside carbon within soot \citep{Alexander12}.  Thus, other chemical solutions for hazes are possible.  Regardless, methane will be present in these systems in abundance.

\begin{figure}
    \centering
    \includegraphics[scale=0.785]{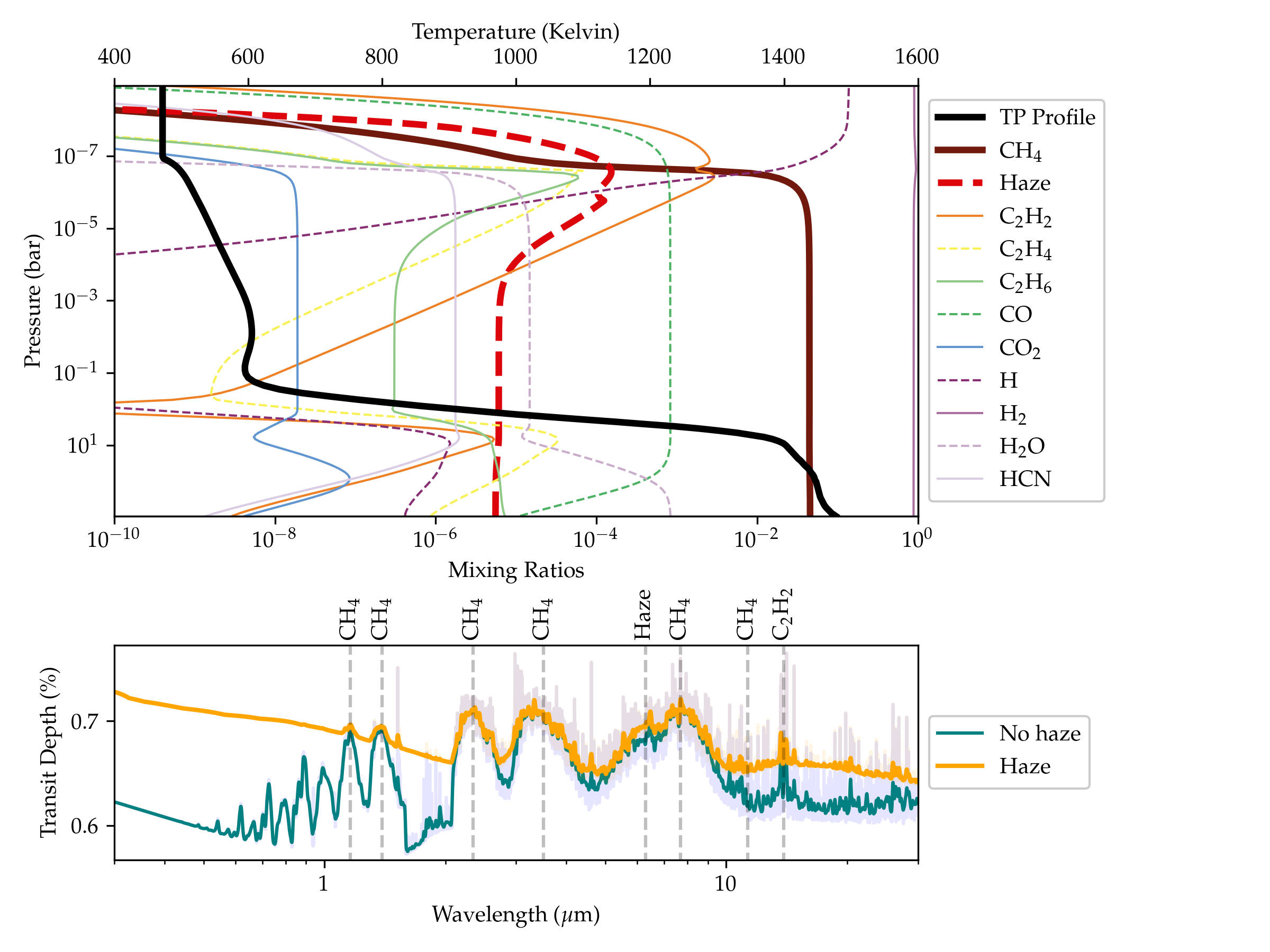}
    \caption{\bf {\em Top:} Abundance and temperature-pressure profiles for a 3 $M_\oplus$, 600 K equilibrium temperature planet, with 0.1\% soot, orbiting an M-dwarf star.  Even under the harsh UV irradiation environment of the host star, high abundances of methane (thick brown line) persist to high altitude and low pressure.  A combination of methane photolysis in the upper atmosphere, vertical mixing, and thermochemistry give rise to significant quantities of higher-order hydrocarbons such as C$_2$H$_2$, C$_2$H$_4$, and C$_2$H$_6$,  Subsequent photolysis and polymerization reactions result in the formation of hydrocarbon haze (thick dashed red line). {\em Bottom:}  The resulting transmission spectrum (orange line) of this planet is shaped considerably by haze with some strong methane features permeating through the haze at longer wavelengths.  The teal line shows the transmission spectrum of the same planet but with the haze opacity artificially removed, while the transparent colored line shows these data at a spectral resolution of R=1,000.}
   \label{fig:transmission} 
\end{figure}

The baseline haze model is discussed in  Appendix \ref{meth:atmosphere}.
We specifically model the planet  at an equilibrium temperature of 600~K placed in orbit around an M-dwarf host star to align its properties with sub-Neptune exoplanet targets that will be observed with JWST during its first year of operations.  The resulting chemical abundance profiles are presented in Fig.\ref{fig:transmission}, which demonstrate that these atmospheres readily produce hazes via hydrocarbon polymerization channels. 

\section{Implications}

\subsection{Current Exoplanet Landscape}

Planets larger than Earth but smaller than Neptune (super-Earths and sub-Neptunes) orbiting close-in to their host stars are the most commonly occurring type of planet that we know of in our galaxy \citep{fulton17}.  Such objects are scheduled for ample observation time with JWST in its first year of operation.  The JWST targets mostly orbit lower-mass stars (i.e., M dwarfs), and the goal is to detect spectral features originating from atmospheric gases and aerosol layers.  With respect to our current line of questioning about incorporation of carbon into these planets at birth, we are most interested in identifying the spectral signatures of carbon-bearing molecules in their atmospheres (e.g.\ CH$_4$) and the hazes themselves \citep[e.g.,][]{ohno20}.  

Muted spectral features have been a common theme in observations of sub-Neptune transmission spectra \citep[e.g.][]{bean10, benneke19, guo20}.  Only in the case of the planet GJ 1214b can the lack of atmospheric absorption be definitively interpreted as aerosol obscuration \citep{Kreidberg14}.  In other cases, degeneracies still exist between high mean molecular weight and aerosol interpretations due to the level of precision of existing data.  For warm sub-Neptunes ($T \lesssim 850$ K), the favored interpretation of muted spectral features has been hydrocarbon hazes, formed from pathways that begin with the photolysis of CH$_4$ \citep{Kempton12, kawashima18, lavvas19}.  Positing that CH$_4$ destruction is the catalyst for haze formation, it follows that we should search for the spectroscopic signatures of this gas (and other hydrocarbon haze ``precursors'' such as HCN, C$_2$H$_2$, C$_2$H$_4$, etc.).  Unfortunately, CH$_4$ has been surprisingly challenging to detect, despite its ample strong spectroscopic features within the wavelength range of existing instruments.  Even planets that are cool enough to host considerable CH$_4$ in their atmospheres via thermochemical equilibrium considerations have not produced detectable features \citep{stevenson10, benneke19, fu22}.  What few observational searches for CH$_4$ that do appear in the literature \citep[e.g.][]{swain08, guilluy19, giacobbe21, bezard22} have been called into question by other works or have not been reproduced.  This ``missing methane” problem could have a number of solutions.  The CH$_4$ could be entirely destroyed by photolysis reactions or chemical quenching, the planets could be intrinsically carbon-poor, or the data quality and detection techniques might simply not be sufficient yet (e.g., inaccurate CH$_4$ line lists at high spectral resolution, or the co-mingling of methane and water vapor bands at moderate to high temperatures). 

\subsection{Predictions for JWST and Mantle Water Content}

 The predicted transmission spectrum of a volatile rich world from 0.3 to 30 $\mu$m is shown in Fig.~\ref{fig:transmission}.  The baseline spectrum (no hazes) exhibits strong methane features.  In the presence of hazes the spectrum is significantly muted, in line with existing observations of sub-Neptune atmospheres \citep{Kreidberg14, Knutson14, Libby-Roberts20}.   These models are for 3 M$_\oplus$ planet, and existing spectroscopic data on atmospheres currently do not detect discernible atmospheric features towards these lower mass planets \citep{dewit18,diamondlowe20,libby22}.  The expectation of our model results is that similar features would be anticipated for slightly more massive planets (e.g., 5 M$_\oplus$).

 We find that even adding comparable amounts of soot and water does not dilute the presence of a rich methane-dominated atmosphere (Fig.~\ref{fig:atmcomp}).   To determine the robustness of this model we vary the water content within our baseline model for a  3 ~M$_\oplus$, 600 K equilibrium temperature planet, with 0.1\% soot and no water. The water content in the mantle of this planet is 0.23 wt\% with an atmospheric water mixing ratio of 8 $\times 10^{-4}$ and a total water column depth of  1.5 $\times 10^{26}$ molecules/cm$^2$.  The atmospheric mixing ratio of water provides the floor value for varying the water content (i.e., 1$\times$ \ce{O}).  This might occur if some material is supplied to the young silicate/soot-rich planet from beyond the water ice line.  To model this we raise the mixing ratio of water by integer units noted in Fig.~\ref{fig:atmcomp}, but maintain oxygen mass balance in the calculation by lowering other carriers uniformly.
 %\footnote{\bf We note that we did not self-consistently account for this additional water vapor in the atmosphere and its potential effect on the physical/chemical profile beyond mass balance.}}
 We otherwise maintain the temperature-pressure profile, eddy diffusion coefficient, and stellar spectrum as for the modeling described in \S~3.2.  This model is not self-consistent as we are not modeling all the geochemical steps described in \S 3.1 and do not account for the potential effect of this additional water vapor on the physical/chemical profile. Rather, we vary the atmospheric water content to capture the impact of additional oxygen in the atmosphere on haze production and approximately simulate scenarios in which additional water is provided from beyond the snowline.  
 
 \begin{figure}
    \centering
    \includegraphics[scale=0.785]{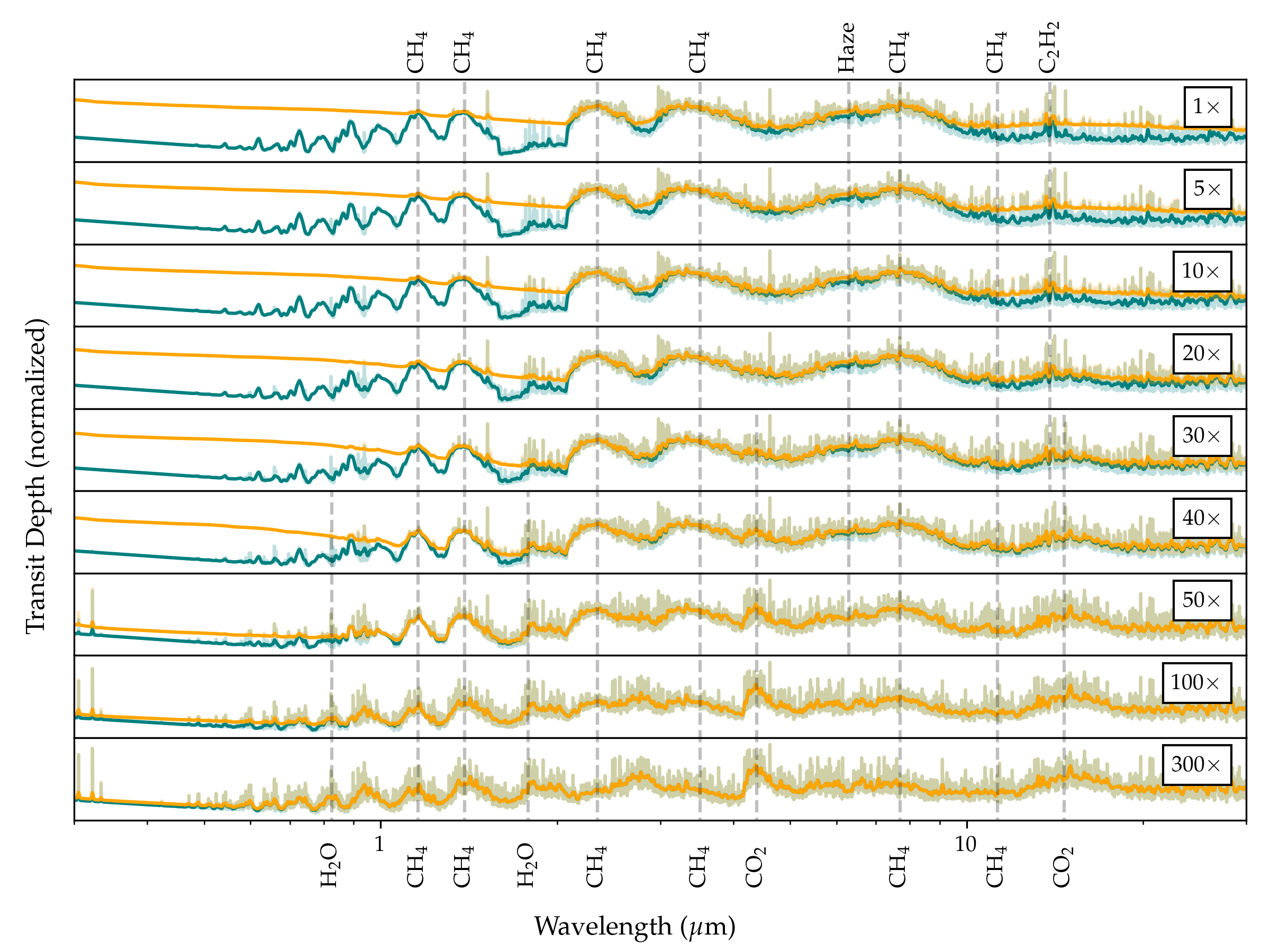}
    %% need  larger fonts for axes labels and annotations
    \caption{\bf Changes in transmission spectra for increasing enhancement of atmospheric oxygen by factors of 1-300$\times$ from top to bottom. The full spectra (orange lines) are compared to ``hazeless'' spectra for which the opacity contribution of haze was removed from the calculation (teal lines), revealing the otherwise muted molecular features of the gas phase species. The transparent colored lines represent the full-resolution transmission spectra output by \texttt{Exo-Transmit} (spectral resolution of $R = 1000$), while the opaque, thick lines are smoothed for ease of visualization. The baseline case (1$\times$) corresponds to Figure \ref{fig:transmission}. The bulk composition of the atmosphere changes with increasing oxidation, and new molecular features of \ce{H2O} and \ce{CO2} become apparent as their abundance increases. Haze dominates the transmission spectrum until oxygen has been added to the atmosphere at $\sim$50$\times$ its baseline value. At an oxygen enhancement of 300$\times$, the production of haze is effectively suppressed and its impact on the transmission spectrum is negligible.   We note that the feature labelled as Haze near 7~$\mu$m is a spectroscopic feature resulting a sharp rise in the extinction coefficient at that wavelength in the Haze model.}
    \label{fig:oxidation}
\end{figure}

 The impact of decreasing haze production on transmission spectra is shown in Fig.~\ref{fig:oxidation}.
 Our atmospheric modeling suggests that substantial \ce{H2O} incorporation during planet formation that translates to additional atmospheric oxygen exceeding a 40-fold increase, such as what would be expected for planets born beyond the snow line, is needed to throttle haze production in this model of a soot-rich planet.
 Overall, we find that these results indicate high atmospheric methane concentration, ample haze production, and transmission spectra dominated by haze and methane features are robust even when considerable excess oxygen is added to the atmosphere from water incorporation at the time of formation. 
 
\subsection{Volatile Rich Worlds}

Carbon-rich planets have been discussed previously \citep{Seager07} 
and 55 Cancri e has been suggested to have a carbon-rich interior \citep{Madhusudhan12}. Much of the focus of carbon-rich planets is within systems where the stars and protoplanetary disks have overall elemental C/O $>$ 1.  Under these assumptions, \citet{Bond10a} find that within dynamical simulations, under equilibrium chemical conditions, bulk planets can form with tens of wt\% of carbon with carbon provided in the form of graphite, TiC, and SiC.  Our hypothesis differs somewhat as organic-rich  soot is the primary source of carbon in our solar system (and likely others) and the species that comprise this material are not products of equilibrium condensation \citep{Li21}.  In our model the bulk system (i.e. the star and disk) has C/O $<$ 1 but the resulting atmospheric composition develops C/O $>$ 1 because of outgassing from the reduced carbon-rich and water-poor mantle.

High carbon abundances are also expected within exoplanets formed through pebble accretion as opposed to the planetesimal accretion considered by \citet{Bond10a}.  In this mode of growth, small ($<$ 1 m) solids are readily accreted by growing embryos as they drift towards the star due to aerodynamic interactions with the surrounding protoplanetary disk gas \citep[e.g.][]{Lambrechts2019}.  These pebbles would lose volatiles as they crossed various snow lines, meaning that those planets growing inside of the snow line but exterior to the soot line would readily accrete carbon-rich materials provided this feedstock is not limited by large, Jovian-mass planets growing further out in the disk \citep{Mulders2021}.  As shown in Fig.~\ref{fig:exo}, such planets are likely among the population of known exoplanets and will be targeted for atmospheric characterization by JWST in the coming years.

%The majority of exoplanets were born close to their stars in systems that lack a Jovian planet at large distances, allowing the super-Earths to form as they derive their mass from pebbles that drift inwards from further out in the disk \citep{Lambrechts2019}.  This mode of growth naturally explains why planet occurrence rates at short orbital periods are higher among lower mass stars \citep{Mulders2021}.These pebbles would lose volatiles as they crossed various snow lines, meaning that soot-rich rocky worlds must exist as they would have formed interior of the water snow line and exterior to the soot line.  Not only must they exist, as shown in Fig.~\ref{fig:exo}, they are likely among the population of known exoplanets and will be targeted for atmospheric characterization by JWST in the coming years. 
%Thus we predict that hazy atmospheres will be common among the early JWST data release.

Our work has the most direct relevance towards somewhat more massive planets with hydrogen rich envelopes.  Flattened transmission spectra have already been detected towards gas-rich small planets  and have been posited as due to photochemical hazes \citep{Kempton12}. These hazes could readily be a by-product of birth between the soot and ice lines. Such hazes, and the methane that drives their formation, are detectable via JWST transit spectroscopy, as demonstrated here, especially around stars lower in mass (and therefore size) than the Sun.  Thus, the presence or lack of hazes in the atmospheres of super-Earths or sub-Neptunes may allow us to discern whether they formed in-situ from local materials or closer to the snow line and then migrated inward. For planets comparable in mass to the Earth, the overall evolution needs to be modeled in the future, but presents exciting new avenues for gains in our understanding of planets with significant volatile inventories.

\begin{acknowledgments}
This research comes from an interdisciplinary collaboration initiated by the Integrated NSF Support Promoting Interdisciplinary Research and Education Program through Grant AST1344133. Additional funding has been provided by National Aeronautics and Space Administration Grants 80NSSC19K0959 (to M.M.H.), XRP NNX16AB48G (to G.A.B.), XRP 80NSSC20K0259 (to E.A.B. and F.J.C.), and 80GSFC21M0002 (to S.T.B); and National Science Foundation Grant AAG 2009095 (to E.M.-R.K. and supporting D.J.T.). S.T.B also received support from the GSFC Sellers Exoplanet Environments Collaboration (SEEC), which is funded in part by the NASA Planetary Science Division's Internal Scientist Funding Model.   This project is supported, in part, by funding from Two Sigma Investments, LP to EAB. Any opinions, findings,
and conclusions or recommendations expressed in this material are those of the authors and do not
necessarily reflects the views of Two Sigma Investments, LP.
\end{acknowledgments}

\facilities{JWST}


\software{\texttt{HELIOS} \citep{Malik17,Malik19}, \texttt{Exo-Transmit}  \citep{Kempton17, Teal22}
          }

%\noindent {\bf Fig. 1.} Please do not use figure environments to setup your figures in the final (post-peer-review) draft, do not include graphics in your source code, and do not cite figures in the text using \LaTeX\ \verb+\ref+ commands.  Instead, simply refer to the figure numbers in the text per {\it Science\/} style, and include the list of captions at the end of the document, coded as ordinary paragraphs as shown in the \texttt{scifile.tex} template file.  Your actual figure files should be submitted separately.



\appendix

\section{Mantle/Atmospheric Equilibrium Model \label{meth:mantle_atm}}

Initial calculations begin with different fractions of C-H-O soot (C$_{100}$H$_{77}$O$_{15}$) and silicate, assuming that the silicate initially contains 16 wt. \% FeO, a comparatively oxidized assumption similar to the mantle of Mars.  

We consider a planet with Earth-like proportions of core (33 \% by mass) and silicate (67 \%) and for the condensed portion of the planet, the mass-radius relationship is given by the parameterization a+b M$_\oplus$+c ln (M$_\oplus$), where a= 0.9868, b=0.0231, c=0.2599 are empirical coefficients taken from mass-radius relationships of Earth-like planets from \citet{Santerne18,Zeng19}. This relationship allows for the calculation of the gravitational acceleration at the surface for each planetary mass.

For equilibration between molten silicate and overlying atmosphere, we employ the thermodynamic outgassing model presented in \citet{Gaillard22} (https://calcul-isto.cnrs-orleans.fr/apps/planet/).  This model calculates partial pressures of outgassed species (H$_{2}$, H$_{2}$O, CO$_{2}$, CO, CH$_{4}$) based on assumed total mantle mass, volatile content, temperature (1773 K for our calculations), and oxygen fugacity, but the critical values, passed to the atmospheric calculations described below, are the total elemental masses of outgassed elements (principally C-H-O). 

 In the calculation of  \citet{Gaillard22}, the calculated masses of silicate outgassed volatiles do not conserve oxygen mass balance.  This is because volatile inputs are taken only as oxidized species (H$_{2}$O, CO$_{2}$), but output as both oxidized and reduced C-H-O species.  This required several adjustments.  First, input of reduced volatiles, such as hydrogen gas or "soot" required adjustments according to reactions such as  H$_{2}$ + FeO = Fe + H$_{2}$O.
 
 Second, because the \citet{Gaillard22} calculator assumes that the silicate is effectively an infinite reservoir, leaving oxygen fugacity unchanged even as oxidized input species are converted to a combination of oxidized and reduced species, we took an iterative approach.  After each iterative step, we recalculated the concentration of FeO in the silicate by enforcing O mass balance, and then calculated f$_{O_2}$ based on an empirical curve derived from \citet{Frost08}: $\Delta$IW=0.8763$\cdot\ln$(FeO)-3.80 (IW = Iron-Wustite buffer), where FeO is in units of wt.$\%$.  Oxygen fugacities were bounded at a minimum of $\Delta$IW=-6, below which melt FeO is effectively zero, making mass balance ineffective.  Convergence was accepted when resulting f$_{O_2}$ (oxygen fugacity) differed by less than 0.03 log units from the previous iteration, and generally 4-5 iterative steps were required.

The temperature for the principal calculations was selected as 1773 K because these are conditions close to the experimental constraints on volatile solubilities in silicate liquid employed by the calculator.  Greater temperatures are expected at the surfaces of magma oceans, particularly for larger planets, and this will affect volatile speciation.  For example, at higher temperature, CO and H$_{2}$ are favored relative to CH$_{4}$ and H$_{2}$O.  This effect is not so consequential for the purposes of the present calculation, as the information from the outgassing calculation that is passed to the atmospheric calculations described below is that of elemental abundances, rather than gaseous species.  To illustrate the temperature sensitivity of our calculations, we include one calculation at 2773 K for the case of 0.1$\%$ soot and 1 wt.$\%$ H$_{2}$O.  As shown in Table A.1, the resulting elemental abundances of the initial outgassed atmosphere are not consequentially different for the same bulk composition at 1773 K.

\setcounter{table}{0}
\renewcommand{\thetable}{A\arabic{table}}
\begin{center}
    \begin{table}
        \caption{Model Assumptions and Base Atmosphere  Properties/Composition with variable Temperature}
        \centering
    \begin{tabular}{cccrrrrrrrr}
    \hline
    \multicolumn{1}{c}{\% Soot$^a$} &
    \multicolumn{1}{c}{\% H$_2$O$^a$} & \multicolumn{1}{c}{M$_p$} &
 \multicolumn{1}{c}{M$_{p,soot}$} &
    \multicolumn{1}{c}{M$_{p,H_2}$} &
    \multicolumn{1}{c}{P} &
    \multicolumn{1}{c}{T} &   \multicolumn{1}{c}{log$_{10}$(f$_{\rm O_2}$)} & 
        \multicolumn{3}{c}{Element Fractions}\\
    \multicolumn{2}{c}{(by mass)} &
    \multicolumn{3}{c}{(M$_\oplus$)} & \multicolumn{1}{c}{(MPa)} & \multicolumn{1}{c}{(K)} & \multicolumn{1}{c}{}  &\multicolumn{1}{c}{H}&\multicolumn{1}{c}{O}&\multicolumn{1}{c}{C}\\\hline\hline
%%soot  %h2o Planet mass (Earth Masses)	P (MPa)	N2	H2O	H2	CO2	CO	CH4

0.1 &1.0 & 0.3& 0.003 & 8.7$\times 10^{-8}$ & 109.3 & 1773 & -10.36 & 0.668	&0.165 &	0.167\\
0.1 &	1.0	&0.3	&0.003&	8.7$\times 10^{-8}$&105.9& 2773 & -4.85 &	0.666&	0.167&	0.167\\
0.1 &1.0 &  1.0& 0.01 & 1.5$\times 10^{-4}$ & 188.2 & 1773 & -10.43 & 0.679	&0.135	&0.186\\
0.1 &	1.0	&  1.0 &	0.01	&1.5$\times 10^{-4}$ &	186.0& 2773 &	-4.85 & 0.674 &	0.137 &	0.189\\
0.1 &1.0 &  3.00& 0.03 & 1.4$\times 10^{-2}$ &1120.0 & 1773 & -14.84 & 0.983	&0.000	&0.017\\
0.1 &	1.0	&3.0	&0.03 &	1.4$\times 10^{-2}$	&1120.0	&2773 & -9.07 &	0.983	&0.000&	0.017\\ \hline
\multicolumn{10}{l}{$^a$Percent mass added to M$_p$.}
    \end{tabular}
    \label{tab:geochemT}
\end{table}
\end{center}

A rigorous treatment of volatile reservoirs developed during planetary differentiation would include sequestration of C and H in the core, as both elements are siderophile \citep{Hirose2019, Li21}. Developing a model that includes partitioning of C and H into the core would require addressing the processes of accretion and core segregation that precede the initial conditions of our model, which begins with a full-formed planet from which the core has already segregated.  However, the absence of this reservoir does not have a strong effect on the modeling we present because the amount of soot in each calculation is a free variable.  For a fixed supply of volatile materials (i.e., soot, ice) to a planet of a given mass, the capture of volatiles by the core would reduce their masses residing in the mantle and atmosphere, thereby ultimately diminishing the total atmospheric pressure. Core segregation does not change the speciation of carbon in the mantle, which occurs as accessory phases including diamond, graphite, iron carbide, and C-H-O fluid \citep[e.g.,][]{Frost08}. Therefore, for the purposes of investigating methane outgassing and haze production, a model that incorporates core segregation and a greater amount of accreted soot would give essentially the same results as a model with no core segregation and a smaller amount of soot. 

\section{Modeling the Observable Atmosphere}
\label{meth:atmosphere}
To model the composition of the portion of the atmosphere that would be observable via spectroscopic techniques with JWST, we apply two different methodologies.  The first is a chemical equilibrium calculation of atmospheric abundances as a function of atmospheric pressure.  The second is a chemical kinetics calculation that includes photolysis reactions and photochemical production of important hydrocarbon haze precursors, and thus haze.  

\begin{figure}
    \centering
    \includegraphics[scale=0.6]{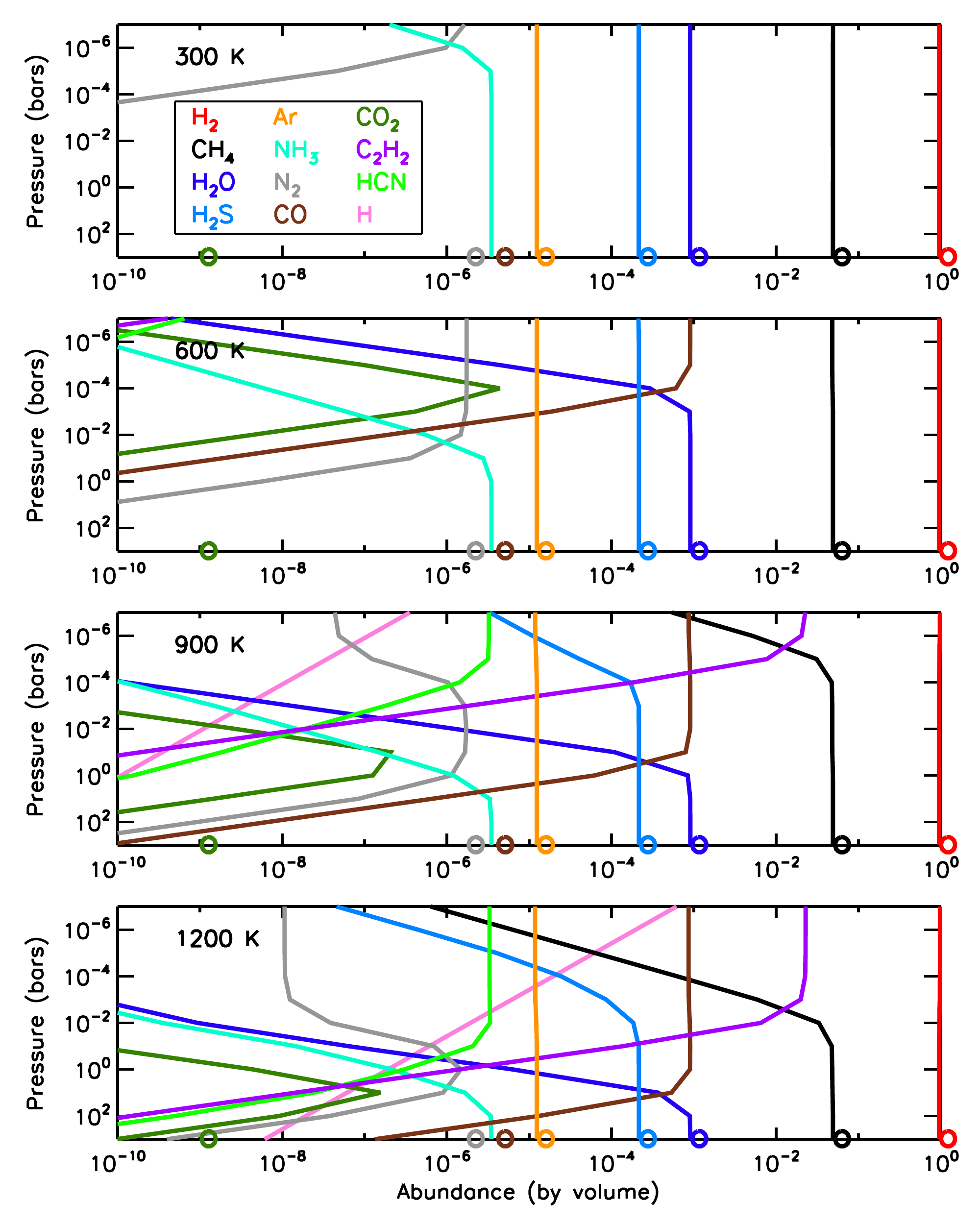}
    \caption{\bf Equilibrium chemical abundances for a 3 $M_{\oplus}$ planet with 0.1\% soot for isothermal atmospheres at different temperatures. Across the entire temperature range considered, methane (black) or acetylene (violet) persists at high concentration through the observable portion of the atmosphere ($\sim 1$ mbar in transmission spectroscopy).  As shown in our photochemical modeling, methane photolysis brings about the formation of organic hazes, and acetylene is an even more direct haze precursor.  Circles at the bottom of the plot panels show the bottom-of-atmosphere chemical abundances from our mantle/atmosphere equilibrium model.  The plotted abundance profiles would exactly match the circle symbols at the bottom-of-atmosphere temperature and pressure values derived in the mantle/atmosphere equilibrium model (Table~\ref{tab:geochem}).}
    \label{fig:equilabun}
\end{figure}

For the chemical equilibrium calculation, we start from the surface composition at the atmosphere-mantle boundary  (Table~\ref{tab:geochem}), calculated as described in Section~\ref{meth:mantle_atm}.  We then derive the underlying \textit{elemental} abundances (i.e.,  H$_2$O $\rightarrow$ 2 H $+$ 1 O) of H, C, N, O, S, and Ar.  From these abundances, we re-derive thermochemical equilibrium as a function of temperature and pressure using the Gibbs free energy minimization techniques described in \citet{mbarek16}.  We perform our calculations over a pressure range of 1 $\mu$bar -- 100 bar and for temperatures from 300~K to 1200~K for a set of 69 molecules made up of H, C, N, O, and S (and Ar).  An example of the resulting chemical equilibrium abundances vs.\ pressure is shown in Fig.\ref{fig:equilabun}.  We additionally show the chemical equilibrium abundances at a pressure of 1 mbar (approximately the pressure level probed in transmission spectroscopy) for planets across our entire model domain in Fig.~\ref{fig:atmcomp} in the main text.

For the chemical kinetics modeling, we first must generate realistic temperature-pressure (T-P) profiles for the atmospheres in question.  (This step is unnecessary for the chemical equilibrium modeling, described above, because in that case the chemical composition depends uniquely on the local temperature and pressure of the gas, rather than the full vertical T-P profile.)  We use the open-source \texttt{HELIOS}\footnote{\url{https://github.com/exoclime/HELIOS}} code \citep{Malik17,Malik19} to calculate temperature-pressure profiles in radiative convective equilibrium.  We generate T-P profiles for the 3 $M_\oplus$ planet, which for reasons already discussed in the text is the scenario for which we believe our modeled atmospheres are most representative of the evolved planets that will typically be observed with JWST.  We model planets with equilibrium temperatures of 600, 900, and 1200 K (we focus on the 600~K model in the main text), set by selecting the planet's orbital semi-major axis assuming zero albedo and fully efficient day-night heat redistribution. The pressure at the bottom of the atmosphere is set to 10$^{3}$ bar. The host star properties and spectrum are selected to match the M-dwarf star GJ 876 ($T_{eff} = 3300$ K, $R_{\star} = 0.367 R_{\odot}$) as representative of a typical system that would be observed with JWST.

The resulting \texttt{HELIOS} T-P profiles are then passed into a chemical kinetics code to calculate atmospheric abundances of gas-phase species and hydrocarbon haze as a function of altitude. As the impact of photodissociation is particularly pronounced at low pressures beyond the pressure cut-offs commonly used in radiative transfer models (here: $10^{-7}$ bar), we extrapolate the \texttt{HELIOS} T-P profiles as isothermal to $10^{-9}$ bar.  We use the version of the \texttt{Atmos} photochemistry code described in \citet{Harman22}, with the addition of carbon-bearing species and chemical reactions up to \ce{C}-4 (\ce{C3H2}, \ce{C3H3}, \ce{C3H4}, \ce{C4H2}, \ce{C4H3}, \ce{C4H5}) and nitrogen-bearing species and reactions (\ce{N2}, \ce{N}, \ce{NH}, \ce{NH2}, \ce{NH3}, \ce{N2H}, \ce{N2H2}, \ce{N2H3}, \ce{CN}, \ce{NCO}, \ce{HCN}, \ce{HNO}, \ce{HNCO}, \ce{NO}, \ce{H2CN}, \ce{HC3N},  \ce{C2H3CN}, \ce{CH2NH}, \ce{CH2NH2}, \ce{CH3NH2}, \ce{CH2CN}, \ce{CH3CN}) sourced from \citet{Tsai_2021}.  We additionally account for the formation of organic haze using the fractal haze model from \citet{Wolf_2010,Arney16,Arney17} adapted for an H$_2$-dominated atmosphere \citep{Parmentier_2013}.  Haze formation is primarily initiated by CH$_4$ photolysis, which catalyzes the formation of complex organic molecules in the atmosphere.  Our chemical network cannot capture the full complexity of reactions occurring among all of these high-order hydrocarbon molecules.  We instead follow a common practice of selecting lower-order haze ``precursor" species from our chemical network that are formed high up in the atmosphere. For the current work we select polyacetylene (C$_{2n}$H$_2$) [e.g. \citep{allen1980,WILSON_2003,Lavvas_2008_1}] and allene (\ce{CH2CCH2}) polymerization \citep{Pavlov2001a} pathways, both proceeding through reactions with the ethynyl radical \ce{C2H}, and a nitrogen bearing co-polymer pathway based on cyanoacetylene \ce{HC3N} \citep{Krasnopolsky_2009,Lavvas_2008_1} for haze production:
\begin{equation} \label{eq:haze pw1}
  \ce{  C4H2 + C2H  -> Polymer + H} 
\end{equation}
\begin{equation} \label{eq:haze pw2}
    \ce{CH2CCH2 + C2H -> Polymer + H} 
\end{equation}
\begin{equation} \label{eq:haze pw3}
    \ce{HC3N + C4H3 -> }\ce{Co}\textnormal{-}\ce{Polymer} 
\end{equation}
We assume a 100\% conversion efficiency into haze.  Once hazes form in the photochemistry model they scatter and absorb incoming UV photons, which ultimately self-regulates the formation of additional haze. Aerosol particles form as Mie scatterers that grow and coagulate into fractal aggregate particles composed of monomers of a fixed size of 50 nm. Haze optical properties for spherical and fractal aggregate particles were calculated with the mean field approximation model described in  \citet{RANNOU1999} and \citet{Botet1997} assuming Titan tholin complex refractive indices from \citet{khare84}.  The irradiating host star is again selected to be GJ 876, using its UV spectrum from the MUSCLES catalog \citep{France16, https://doi.org/10.17909/t9dg6f}\footnote{\url{https://archive.stsci.edu/prepds/muscles/}}.  We assume a uniform Eddy diffusion coefficient of $\textnormal{K}_{zz}=6 \times 10^{8} $ cm$^{2}$  s$^{-1} $, similar in range as previous studies \citep{Kawashima_2018,Tsai_2021,Harman22}. %(XXX kawashima actually uses 1e7 and tries several, Tsai uses a range, sonny uses 1e10).
While the choice of $\textnormal{K}_{zz}$ influences particle coagulation and atmospheric mixing, we forgo a detailed discussion and note that all atmospheric models we generated produced significant amounts of haze for the complete range of  $\textnormal{K}_{zz}$ values  we tested ($5 \times 10^{8} -  5 \times 10^{10}$ cm$^{2}$  s$^{-1} $). The atomic composition determined above in the chemical equilibrium modeling was scaled to preserve the relative abundance ratios while introducing a solar metallicity abundance of \ce{He}, which \texttt{Atmos}  uses as a (required) non-reactive filler gas. The planet's gravity at the 10$^{3}$ bar level and radius were set to 1481.86 cm s$^{-2}$ and 1.41 R$_{\oplus}$, respectively.

Finally, we model the transmission spectra of the resulting atmospheres.  For this we use the \texttt{Exo-Transmit} code \citep{Kempton17}, as modified in \citet{Teal22}, to generate transmission spectra from the vertical abundance profiles output by the chemical kinetics code.  Haze opacities are included in this version of \texttt{Exo-Transmit}, which depend on the haze particle radius.  We use an identical set of hydrocarbon haze optical properties for all haze particles in the atmosphere, regardless of which of the three precursor formation pathways generated the haze. In Figure \ref{fig:transmission} of the main text, we show versions of the transmission spectra with the haze opacity included and removed, emphasizing the impact of hazes on muting/obscuring spectral features.

%\section{Influence of uncertainties in \ce{H2O} incorporation on photochemical haze production rates}
%\label{meth:H2Oincrease}
%Additional \ce{H2O} incorporation during planet formation may render haze production ineffective, as the atmospheric photolysis of oxygen-bearing species such as \ce{H2O} and \ce{CO2} releases oxidizing radicals ( O($^1$D),  O($^3$P), \ce{OH}) that can interrupt the polymerization of hydrocarbon chains.  
%We test the robustness of our chemical kinetics model results against significant reduction in C/O through additional \ce{H2O} by gradually increasing the stoichiometric ratio of \ce{O} (enhancement of \ce{O} ranging from 1-300$\times$ the baseline \ce{O} content) in the atmosphere.  . %\textcolor{red}{We forgo a mantle/atmospheric equilibrium modelling here. Same TP, Eddy}  %I need to justify why we do not use the full mantle/atmospheric equilibrium model. Why is it sufficient to just assume we have more O, and not take into account how C and H would be affected?

\begin{figure}
    \centering
    \includegraphics[scale=0.12]{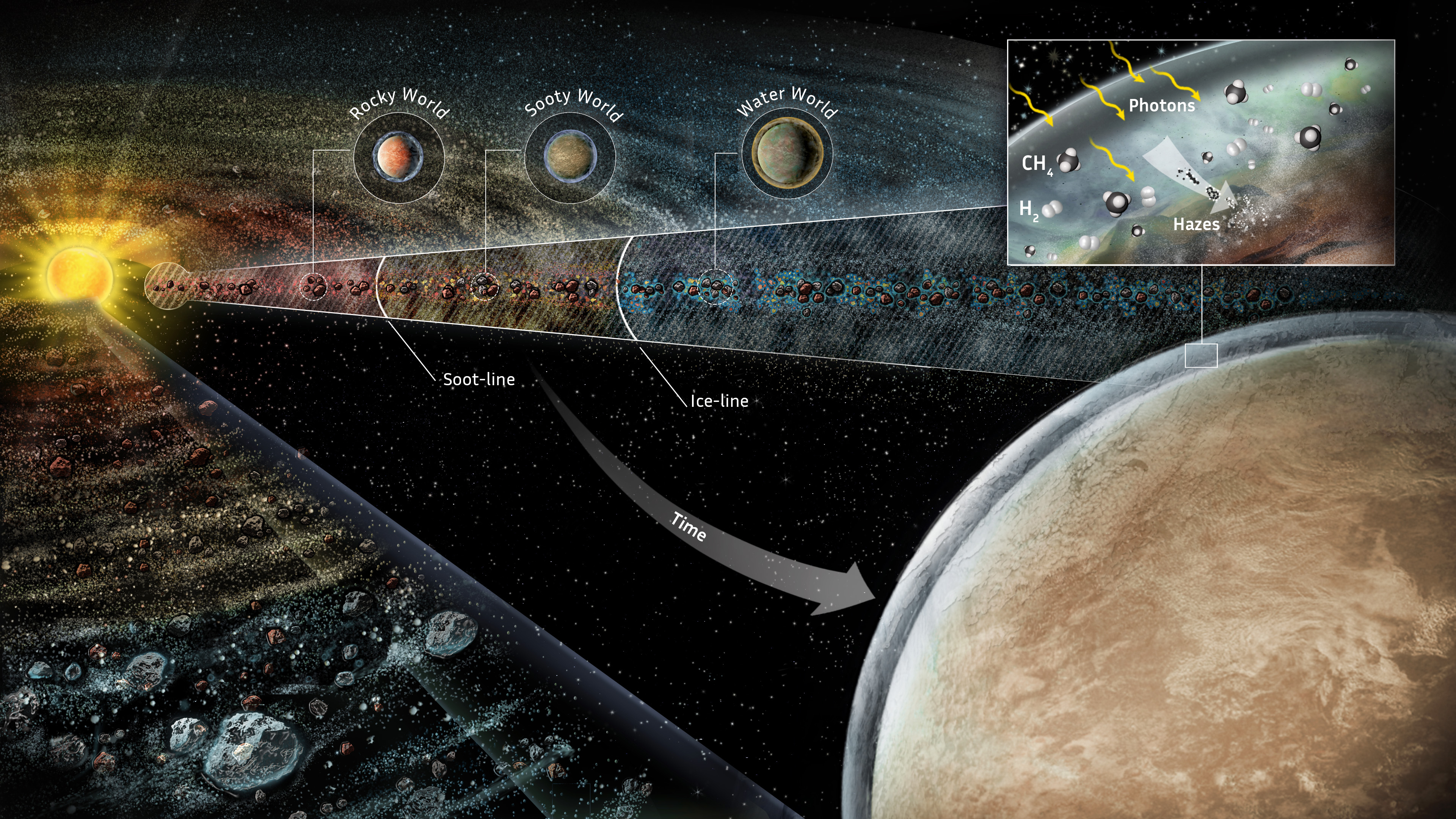}
    \caption{\bf Schematic of disk chemical structure.  We focus on objects built from the abundant solids -- silicates, refractory organics, and water ice.
    The temperature declines with distance from the star with key chemical transition points being the soot line, where refractory organics sublimate into gas, and the water ice line, which denotes the water ice 
   sublimation front. A planet that receives most of its material interior to the soot line will be silicate-dominated.  A planet born with materials from within the soot and ice lines will be a soot-rich silicate body, while planets formed beyond the ice line will be comprised of all three major solid state components.  A planet that receives most of its initial materials from between the soot and ice line will have a soot-rich mantle. At later time after a planet is assembled (time arrow to planet with atmosphere), the combination of carbon outgassing from the mantle and, for more massive planets, the presence of hydrogen envelopes will lead to copious CH$_4$ (and other hydrocarbon) formation.  This methane persists to altitude and, when exposed to stellar UV light, facilitates haze formation as shown in the call out box. {\em Image Credit: Ari Gea/SayoStudio}}
   \label{fig:schematic} 
\end{figure}

Fig.~\ref{fig:schematic} is provided as a bonus figure for ArXiv version as a schematic of the various key chemical transitions in the inner disk including the soot and water ice lines.

\bibliography{z}{}

\begin{thebibliography}{}
\expandafter\ifx\csname natexlab\endcsname\relax\def\natexlab#1{#1}\fi
\providecommand{\url}[1]{\href{#1}{#1}}
\providecommand{\dodoi}[1]{doi:~\href{http://doi.org/#1}{\nolinkurl{#1}}}
\providecommand{\doeprint}[1]{\href{http://ascl.net/#1}{\nolinkurl{http://ascl.net/#1}}}
\providecommand{\doarXiv}[1]{\href{https://arxiv.org/abs/#1}{\nolinkurl{https://arxiv.org/abs/#1}}}

\bibitem[{{Alexander} {et~al.}(2012){Alexander}, {Bowden}, {Fogel}, {Howard},
  {Herd}, \& {Nittler}}]{Alexander12}
{Alexander}, C.~M.~O.~., {Bowden}, R., {Fogel}, M.~L., {et~al.} 2012, Science,
  337, 721, \dodoi{10.1126/science.1223474}

\bibitem[{{Alexander} {et~al.}(2013){Alexander}, {Howard}, {Bowden}, \&
  {Fogel}}]{Alexander13}
{Alexander}, C.~M.~O.~., {Howard}, K.~T., {Bowden}, R., \& {Fogel}, M.~L. 2013,
  \gca, 123, 244, \dodoi{10.1016/j.gca.2013.05.019}

\bibitem[{{Alexander} {et~al.}(2017){Alexander}, {Cody}, {De Gregorio},
  {Nittler}, \& {Stroud}}]{Alexander17}
{Alexander}, C.~M.~O.~D., {Cody}, G.~D., {De Gregorio}, B.~T., {Nittler},
  L.~R., \& {Stroud}, R.~M. 2017, Chemie der Erde / Geochemistry, 77, 227,
  \dodoi{10.1016/j.chemer.2017.01.007}

\bibitem[{{Allen} {et~al.}(1980){Allen}, {Yung}, \& {Pinto}}]{allen1980}
{Allen}, M., {Yung}, Y.~L., \& {Pinto}, J.~P. 1980, \apjl, 242, L125,
  \dodoi{10.1086/183416}

\bibitem[{{Andrews} {et~al.}(2016){Andrews}, {Wilner}, {Zhu}, {Birnstiel},
  {Carpenter}, {P{\'e}rez}, {Bai}, {{\"O}berg}, {Hughes}, {Isella}, \&
  {Ricci}}]{Andrews16}
{Andrews}, S.~M., {Wilner}, D.~J., {Zhu}, Z., {et~al.} 2016, \apjl, 820, L40,
  \dodoi{10.3847/2041-8205/820/2/L40}

\bibitem[{{Arney} {et~al.}(2016){Arney}, {Domagal-Goldman}, {Meadows}, {Wolf},
  {Schwieterman}, {Charnay}, {Claire}, {H{\'e}brard}, \& {Trainer}}]{Arney16}
{Arney}, G., {Domagal-Goldman}, S.~D., {Meadows}, V.~S., {et~al.} 2016,
  Astrobiology, 16, 873, \dodoi{10.1089/ast.2015.1422}

\bibitem[{{Arney} {et~al.}(2017){Arney}, {Meadows}, {Domagal-Goldman},
  {Deming}, {Robinson}, {Tovar}, {Wolf}, \& {Schwieterman}}]{Arney17}
{Arney}, G.~N., {Meadows}, V.~S., {Domagal-Goldman}, S.~D., {et~al.} 2017,
  \apj, 836, 49, \dodoi{10.3847/1538-4357/836/1/49}

\bibitem[{{Bardyn} {et~al.}(2017){Bardyn}, {Baklouti}, {Cottin}, {Fray},
  {Briois}, {Paquette}, {Stenzel}, {Engrand}, {Fischer}, {Hornung}, {Isnard},
  {Langevin}, {Lehto}, {Le Roy}, {Ligier}, {Merouane}, {Modica},
  {Orthous-Daunay}, {Ryn{\"o}}, {Schulz}, {Sil{\'e}n}, {Thirkell}, {Varmuza},
  {Zaprudin}, {Kissel}, \& {Hilchenbach}}]{Bardyn17}
{Bardyn}, A., {Baklouti}, D., {Cottin}, H., {et~al.} 2017, \mnras, 469, S712,
  \dodoi{10.1093/mnras/stx2640}

\bibitem[{{Batygin} \& {Morbidelli}(2023)}]{Batygin23}
{Batygin}, K., \& {Morbidelli}, A. 2023, Nature Astronomy,
  \dodoi{10.1038/s41550-022-01850-5}

\bibitem[{{Bean} {et~al.}(2010){Bean}, {Miller-Ricci Kempton}, \&
  {Homeier}}]{bean10}
{Bean}, J.~L., {Miller-Ricci Kempton}, E., \& {Homeier}, D. 2010, \nat, 468,
  669, \dodoi{10.1038/nature09596}

\bibitem[{{Benneke} {et~al.}(2019){Benneke}, {Wong}, {Piaulet}, {Knutson},
  {Lothringer}, {Morley}, {Crossfield}, {Gao}, {Greene}, {Dressing},
  {Dragomir}, {Howard}, {McCullough}, {Kempton}, {Fortney}, \&
  {Fraine}}]{benneke19}
{Benneke}, B., {Wong}, I., {Piaulet}, C., {et~al.} 2019, \apjl, 887, L14,
  \dodoi{10.3847/2041-8213/ab59dc}

\bibitem[{{Bergin} {et~al.}(2015){Bergin}, {Blake}, {Ciesla}, {Hirschmann}, \&
  {Li}}]{Bergin15}
{Bergin}, E.~A., {Blake}, G.~A., {Ciesla}, F., {Hirschmann}, M.~M., \& {Li}, J.
  2015, Proceedings of the National Academy of Science, 112, 8965,
  \dodoi{10.1073/pnas.1500954112}

\bibitem[{{B{\'e}zard} {et~al.}(2022){B{\'e}zard}, {Charnay}, \&
  {Blain}}]{bezard22}
{B{\'e}zard}, B., {Charnay}, B., \& {Blain}, D. 2022, Nature Astronomy, 6, 537,
  \dodoi{10.1038/s41550-022-01678-z}

\bibitem[{{Birnstiel} {et~al.}(2018){Birnstiel}, {Dullemond}, {Zhu}, {Andrews},
  {Bai}, {Wilner}, {Carpenter}, {Huang}, {Isella}, {Benisty}, {P{\'e}rez}, \&
  {Zhang}}]{Birnstiel18}
{Birnstiel}, T., {Dullemond}, C.~P., {Zhu}, Z., {et~al.} 2018, \apjl, 869, L45,
  \dodoi{10.3847/2041-8213/aaf743}

\bibitem[{{Bond} {et~al.}(2010){Bond}, {O'Brien}, \& {Lauretta}}]{Bond10a}
{Bond}, J.~C., {O'Brien}, D.~P., \& {Lauretta}, D.~S. 2010, \apj, 715, 1050,
  \dodoi{10.1088/0004-637X/715/2/1050}

\bibitem[{Botet {et~al.}(1997)Botet, Rannou, \& Cabane}]{Botet1997}
Botet, R., Rannou, P., \& Cabane, M. 1997, Appl. Opt., 36, 8791,
  \dodoi{10.1364/AO.36.008791}

\bibitem[{{Ciesla} {et~al.}(2015){Ciesla}, {Mulders}, {Pascucci}, \&
  {Apai}}]{Ciesla15}
{Ciesla}, F.~J., {Mulders}, G.~D., {Pascucci}, I., \& {Apai}, D. 2015, \apj,
  804, 9, \dodoi{10.1088/0004-637X/804/1/9}

\bibitem[{{Coleman} \& {Nelson}(2014)}]{Coleman14}
{Coleman}, G. A.~L., \& {Nelson}, R.~P. 2014, \mnras, 445, 479,
  \dodoi{10.1093/mnras/stu1715}

\bibitem[{{Crossfield} \& {Kreidberg}(2017)}]{Crossfield17}
{Crossfield}, I. J.~M., \& {Kreidberg}, L. 2017, \aj, 154, 261,
  \dodoi{10.3847/1538-3881/aa9279}

\bibitem[{{D'Alessio} {et~al.}(2001){D'Alessio}, {Calvet}, \&
  {Hartmann}}]{DAlessio01}
{D'Alessio}, P., {Calvet}, N., \& {Hartmann}, L. 2001, \apj, 553, 321,
  \dodoi{10.1086/320655}

\bibitem[{{de Wit} {et~al.}(2018){de Wit}, {Wakeford}, {Lewis}, {Delrez},
  {Gillon}, {Selsis}, {Leconte}, {Demory}, {Bolmont}, {Bourrier}, {Burgasser},
  {Grimm}, {Jehin}, {Lederer}, {Owen}, {Stamenkovi{\'c}}, \&
  {Triaud}}]{dewit18}
{de Wit}, J., {Wakeford}, H.~R., {Lewis}, N.~K., {et~al.} 2018, Nature
  Astronomy, 2, 214, \dodoi{10.1038/s41550-017-0374-z}

\bibitem[{{Diamond-Lowe} {et~al.}(2020){Diamond-Lowe}, {Berta-Thompson},
  {Charbonneau}, {Dittmann}, \& {Kempton}}]{diamondlowe20}
{Diamond-Lowe}, H., {Berta-Thompson}, Z., {Charbonneau}, D., {Dittmann}, J., \&
  {Kempton}, E. M.~R. 2020, \aj, 160, 27, \dodoi{10.3847/1538-3881/ab935f}

\bibitem[{{Dymont} {et~al.}(2022){Dymont}, {Yu}, {Ohno}, {Zhang}, {Fortney},
  {Thorngren}, \& {Dickinson}}]{dymont22}
{Dymont}, A.~H., {Yu}, X., {Ohno}, K., {et~al.} 2022, \apj, 937, 90,
  \dodoi{10.3847/1538-4357/ac7f40}

\bibitem[{France(2016)}]{https://doi.org/10.17909/t9dg6f}
France, K. 2016, Measurements of the Ultraviolet Spectral Characteristics of
  Low-mass Exoplanetary Systems ("MUSCLES"),  STScI/MAST,
  \dodoi{10.17909/T9DG6F}

\bibitem[{{France} {et~al.}(2016){France}, {Loyd}, {Youngblood}, {Brown},
  {Schneider}, {Hawley}, {Froning}, {Linsky}, {Roberge}, {Buccino},
  {Davenport}, {Fontenla}, {Kaltenegger}, {Kowalski}, {Mauas}, {Miguel},
  {Redfield}, {Rugheimer}, {Tian}, {Vieytes}, {Walkowicz}, \&
  {Weisenburger}}]{France16}
{France}, K., {Loyd}, R.~O.~P., {Youngblood}, A., {et~al.} 2016, \apj, 820, 89,
  \dodoi{10.3847/0004-637X/820/2/89}

\bibitem[{{Frost} {et~al.}(2008){Frost}, {Mann}, {Asahara}, \&
  {Rubie}}]{Frost08}
{Frost}, D.~J., {Mann}, U., {Asahara}, Y., \& {Rubie}, D.~C. 2008,
  Philosophical Transactions of the Royal Society of London Series A, 366,
  4315, \dodoi{10.1098/rsta.2008.0147}

\bibitem[{{Fu} {et~al.}(2022){Fu}, {Espinoza}, {Sing}, {Lothringer}, {Dos
  Santos}, {Rustamkulov}, {Deming}, {Kempton}, {Komacek}, {Knutson}, {Albert},
  {Pontoppidan}, {Volk}, \& {Filippazzo}}]{fu22}
{Fu}, G., {Espinoza}, N., {Sing}, D.~K., {et~al.} 2022, \apjl, 940, L35,
  \dodoi{10.3847/2041-8213/ac9977}

\bibitem[{{Fulton} {et~al.}(2017){Fulton}, {Petigura}, {Howard}, {Isaacson},
  {Marcy}, {Cargile}, {Hebb}, {Weiss}, {Johnson}, {Morton}, {Sinukoff},
  {Crossfield}, \& {Hirsch}}]{fulton17}
{Fulton}, B.~J., {Petigura}, E.~A., {Howard}, A.~W., {et~al.} 2017, \aj, 154,
  109, \dodoi{10.3847/1538-3881/aa80eb}

\bibitem[{{Gail} \& {Trieloff}(2017)}]{Gail17}
{Gail}, H.-P., \& {Trieloff}, M. 2017, \aap, 606, A16,
  \dodoi{10.1051/0004-6361/201730480}

\bibitem[{{Gaillard} {et~al.}(2022){Gaillard}, {Bernadou}, {Roskosz},
  {Bouhifd}, {Marrocchi}, {Iacono-Marziano}, {Moreira}, {Scaillet}, \&
  {Rogerie}}]{Gaillard22}
{Gaillard}, F., {Bernadou}, F., {Roskosz}, M., {et~al.} 2022, Earth and
  Planetary Science Letters, 577, 117255, \dodoi{10.1016/j.epsl.2021.117255}

\bibitem[{{Gao} {et~al.}(2017){Gao}, {Marley}, {Zahnle}, {Robinson}, \&
  {Lewis}}]{Gao17}
{Gao}, P., {Marley}, M.~S., {Zahnle}, K., {Robinson}, T.~D., \& {Lewis}, N.~K.
  2017, \aj, 153, 139, \dodoi{10.3847/1538-3881/aa5fab}

\bibitem[{{Gao} {et~al.}(2020){Gao}, {Thorngren}, {Lee}, {Fortney}, {Morley},
  {Wakeford}, {Powell}, {Stevenson}, \& {Zhang}}]{Gao20}
{Gao}, P., {Thorngren}, D.~P., {Lee}, E. K.~H., {et~al.} 2020, Nature
  Astronomy, 4, 951, \dodoi{10.1038/s41550-020-1114-3}

\bibitem[{{Giacobbe} {et~al.}(2021){Giacobbe}, {Brogi}, {Gandhi}, {Cubillos},
  {Bonomo}, {Sozzetti}, {Fossati}, {Guilluy}, {Carleo}, {Rainer},
  {Harutyunyan}, {Borsa}, {Pino}, {Nascimbeni}, {Benatti}, {Biazzo},
  {Bignamini}, {Chubb}, {Claudi}, {Cosentino}, {Covino}, {Damasso}, {Desidera},
  {Fiorenzano}, {Ghedina}, {Lanza}, {Leto}, {Maggio}, {Malavolta}, {Maldonado},
  {Micela}, {Molinari}, {Pagano}, {Pedani}, {Piotto}, {Poretti}, {Scandariato},
  {Yurchenko}, {Fantinel}, {Galli}, {Lodi}, {Sanna}, \& {Tozzi}}]{giacobbe21}
{Giacobbe}, P., {Brogi}, M., {Gandhi}, S., {et~al.} 2021, \nat, 592, 205,
  \dodoi{10.1038/s41586-021-03381-x}

\bibitem[{{Glavin} {et~al.}(2018){Glavin}, {Alexander}, {Aponte}, {Dworkin},
  {Elsila}, \& {Yabuta}}]{Glavin18}
{Glavin}, D.~P., {Alexander}, C.~M.~O., {Aponte}, J.~C., {et~al.} 2018, {The
  origin and evolution of organic matter in carbonaceous chondrites and links
  to their parent bodies}, ed. N.~Abreu (Elsevier), 205--271,
  \dodoi{10.1016/B978-0-12-813325-5.00003-3}

\bibitem[{{Grewal}(2022)}]{Grewal2022}
{Grewal}, D.~S. 2022, \apj, 937, 123, \dodoi{10.3847/1538-4357/ac8eb4}

\bibitem[{{Guilluy} {et~al.}(2019){Guilluy}, {Sozzetti}, {Brogi}, {Bonomo},
  {Giacobbe}, {Claudi}, \& {Benatti}}]{guilluy19}
{Guilluy}, G., {Sozzetti}, A., {Brogi}, M., {et~al.} 2019, \aap, 625, A107,
  \dodoi{10.1051/0004-6361/201834615}

\bibitem[{{Guo} {et~al.}(2020){Guo}, {Crossfield}, {Dragomir}, {Kosiarek},
  {Lothringer}, {Mikal-Evans}, {Rosenthal}, {Benneke}, {Knutson}, {Dalba},
  {Kempton}, {Henry}, {McCullough}, {Barman}, {Blunt}, {Chontos}, {Fortney},
  {Fulton}, {Hirsch}, {Howard}, {Isaacson}, {Matthews}, {Mocnik}, {Morley},
  {Petigura}, \& {Weiss}}]{guo20}
{Guo}, X., {Crossfield}, I. J.~M., {Dragomir}, D., {et~al.} 2020, \aj, 159,
  239, \dodoi{10.3847/1538-3881/ab8815}

\bibitem[{{Harman} {et~al.}(2022){Harman}, {Kopparapu}, {Stef{\'a}nsson},
  {Lin}, {Mahadevan}, {Hedges}, \& {Batalha}}]{Harman22}
{Harman}, C.~E., {Kopparapu}, R.~K., {Stef{\'a}nsson}, G., {et~al.} 2022, \psj,
  3, 45, \dodoi{10.3847/PSJ/ac38ac}

\bibitem[{{Hartmann} {et~al.}(2016){Hartmann}, {Herczeg}, \&
  {Calvet}}]{Hartmann16}
{Hartmann}, L., {Herczeg}, G., \& {Calvet}, N. 2016, \araa, 54, 135,
  \dodoi{10.1146/annurev-astro-081915-023347}

\bibitem[{{He} {et~al.}(2020){He}, {H{\"o}rst}, {Lewis}, {Yu}, {Moses},
  {McGuiggan}, {Marley}, {Kempton}, {Morley}, {Valenti}, \& {Vuitton}}]{He20}
{He}, C., {H{\"o}rst}, S.~M., {Lewis}, N.~K., {et~al.} 2020, \psj, 1, 51,
  \dodoi{10.3847/PSJ/abb1a4}

\bibitem[{{Hirose} {et~al.}(2019){Hirose}, {Tagawa}, {Kuwayama}, {Sinmyo},
  {Morard}, {Ohishi}, \& {Genda}}]{Hirose2019}
{Hirose}, K., {Tagawa}, S., {Kuwayama}, Y., {et~al.} 2019, \grl, 46, 5190,
  \dodoi{10.1029/2019GL082591}

\bibitem[{{Hirschmann} {et~al.}(2021){Hirschmann}, {Bergin}, {Blake}, {Ciesla},
  \& {Li}}]{Hirschmann21}
{Hirschmann}, M.~M., {Bergin}, E.~A., {Blake}, G.~A., {Ciesla}, F.~J., \& {Li},
  J. 2021, \pnas, 118, e2026779118, \dodoi{10.1073/pnas.2026779118}

\bibitem[{{Ida} \& {Lin}(2008)}]{Ida08}
{Ida}, S., \& {Lin}, D.~N.~C. 2008, \apj, 673, 487, \dodoi{10.1086/523754}

\bibitem[{{Izidoro} {et~al.}(2017){Izidoro}, {Ogihara}, {Raymond},
  {Morbidelli}, {Pierens}, {Bitsch}, {Cossou}, \& {Hersant}}]{Izidoro17}
{Izidoro}, A., {Ogihara}, M., {Raymond}, S.~N., {et~al.} 2017, \mnras, 470,
  1750, \dodoi{10.1093/mnras/stx1232}

\bibitem[{{Kawashima} \& {Ikoma}(2018)}]{kawashima18}
{Kawashima}, Y., \& {Ikoma}, M. 2018, \apj, 853, 7,
  \dodoi{10.3847/1538-4357/aaa0c5}

\bibitem[{Kawashima \& Ikoma(2018)}]{Kawashima_2018}
Kawashima, Y., \& Ikoma, M. 2018, The Astrophysical Journal, 853, 7,
  \dodoi{10.3847/1538-4357/aaa0c5}

\bibitem[{{Kempton} {et~al.}(2017){Kempton}, {Lupu}, {Owusu-Asare}, {Slough},
  \& {Cale}}]{Kempton17}
{Kempton}, E. M.~R., {Lupu}, R., {Owusu-Asare}, A., {Slough}, P., \& {Cale}, B.
  2017, \pasp, 129, 044402, \dodoi{10.1088/1538-3873/aa61ef}

\bibitem[{{Khare} {et~al.}(1984){Khare}, {Sagan}, {Arakawa}, {Suits},
  {Callcott}, \& {Williams}}]{khare84}
{Khare}, B.~N., {Sagan}, C., {Arakawa}, E.~T., {et~al.} 1984, \icarus, 60, 127,
  \dodoi{10.1016/0019-1035(84)90142-8}

\bibitem[{{Knutson} {et~al.}(2014){Knutson}, {Benneke}, {Deming}, \&
  {Homeier}}]{Knutson14}
{Knutson}, H.~A., {Benneke}, B., {Deming}, D., \& {Homeier}, D. 2014, \nat,
  505, 66, \dodoi{10.1038/nature12887}

\bibitem[{Krasnopolsky(2009)}]{Krasnopolsky_2009}
Krasnopolsky, V.~A. 2009, Icarus, 201, 226 ,
  \dodoi{https://doi.org/10.1016/j.icarus.2008.12.038}

\bibitem[{{Kreidberg} {et~al.}(2014){Kreidberg}, {Bean}, {D{\'e}sert},
  {Benneke}, {Deming}, {Stevenson}, {Seager}, {Berta-Thompson}, {Seifahrt}, \&
  {Homeier}}]{Kreidberg14}
{Kreidberg}, L., {Bean}, J.~L., {D{\'e}sert}, J.-M., {et~al.} 2014, \nat, 505,
  69, \dodoi{10.1038/nature12888}

\bibitem[{{Kress} {et~al.}(2010){Kress}, {Tielens}, \& {Frenklach}}]{Kress10}
{Kress}, M.~E., {Tielens}, A.~G.~G.~M., \& {Frenklach}, M. 2010, Advances in
  Space Research, 46, 44, \dodoi{10.1016/j.asr.2010.02.004}

\bibitem[{{Krissansen-Totton} {et~al.}(2018){Krissansen-Totton}, {Olson}, \&
  {Catling}}]{Krissansen-Totton18}
{Krissansen-Totton}, J., {Olson}, S., \& {Catling}, D.~C. 2018, Science
  Advances, 4, eaao5747, \dodoi{10.1126/sciadv.aao5747}

\bibitem[{{Kruijer} {et~al.}(2020){Kruijer}, {Kleine}, \& {Borg}}]{Kruijer20}
{Kruijer}, T.~S., {Kleine}, T., \& {Borg}, L.~E. 2020, Nature Astronomy, 4, 32,
  \dodoi{10.1038/s41550-019-0959-9}

\bibitem[{{Lambrechts} {et~al.}(2019){Lambrechts}, {Morbidelli}, {Jacobson},
  {Johansen}, {Bitsch}, {Izidoro}, \& {Raymond}}]{Lambrechts2019}
{Lambrechts}, M., {Morbidelli}, A., {Jacobson}, S.~A., {et~al.} 2019, \aap,
  627, A83, \dodoi{10.1051/0004-6361/201834229}

\bibitem[{Lavvas {et~al.}(2008)Lavvas, Coustenis, \& Vardavas}]{Lavvas_2008_1}
Lavvas, P., Coustenis, A., \& Vardavas, I. 2008, Planetary and Space Science,
  56, 27, \dodoi{https://doi.org/10.1016/j.pss.2007.05.026}

\bibitem[{{Lavvas} {et~al.}(2019){Lavvas}, {Koskinen}, {Steinrueck},
  {Garc{\'\i}a Mu{\~n}oz}, \& {Showman}}]{lavvas19}
{Lavvas}, P., {Koskinen}, T., {Steinrueck}, M.~E., {Garc{\'\i}a Mu{\~n}oz}, A.,
  \& {Showman}, A.~P. 2019, \apj, 878, 118, \dodoi{10.3847/1538-4357/ab204e}

\bibitem[{{Lee} {et~al.}(2014){Lee}, {Chiang}, \& {Ormel}}]{Lee14}
{Lee}, E.~J., {Chiang}, E., \& {Ormel}, C.~W. 2014, \apj, 797, 95,
  \dodoi{10.1088/0004-637X/797/2/95}

\bibitem[{{Li} {et~al.}(2021){Li}, {Bergin}, {Blake}, {Ciesla}, \&
  {Hirschmann}}]{Li21}
{Li}, J., {Bergin}, E.~A., {Blake}, G.~A., {Ciesla}, F.~J., \& {Hirschmann},
  M.~M. 2021, Science Advances, 7, eabd3632, \dodoi{10.1126/sciadv.abd3632}

\bibitem[{{Libby-Roberts} {et~al.}(2020){Libby-Roberts}, {Berta-Thompson},
  {D{\'e}sert}, {Masuda}, {Morley}, {Lopez}, {Deck}, {Fabrycky}, {Fortney},
  {Line}, {Sanchis-Ojeda}, \& {Winn}}]{Libby-Roberts20}
{Libby-Roberts}, J.~E., {Berta-Thompson}, Z.~K., {D{\'e}sert}, J.-M., {et~al.}
  2020, \aj, 159, 57, \dodoi{10.3847/1538-3881/ab5d36}

\bibitem[{{Libby-Roberts} {et~al.}(2022){Libby-Roberts}, {Berta-Thompson},
  {Diamond-Lowe}, {Gully-Santiago}, {Irwin}, {Kempton}, {Rackham},
  {Charbonneau}, {D{\'e}sert}, {Dittmann}, {Hofmann}, {Morley}, \&
  {Newton}}]{libby22}
{Libby-Roberts}, J.~E., {Berta-Thompson}, Z.~K., {Diamond-Lowe}, H., {et~al.}
  2022, \aj, 164, 59, \dodoi{10.3847/1538-3881/ac75de}

\bibitem[{{Lichtenberg} {et~al.}(2019){Lichtenberg}, {Golabek}, {Burn},
  {Meyer}, {Alibert}, {Gerya}, \& {Mordasini}}]{Lichtenberg19}
{Lichtenberg}, T., {Golabek}, G.~J., {Burn}, R., {et~al.} 2019, Nature
  Astronomy, 3, 307, \dodoi{10.1038/s41550-018-0688-5}

\bibitem[{{Long} {et~al.}(2018){Long}, {Pinilla}, {Herczeg}, {Harsono},
  {Dipierro}, {Pascucci}, {Hendler}, {Tazzari}, {Ragusa}, {Salyk}, {Edwards},
  {Lodato}, {van de Plas}, {Johnstone}, {Liu}, {Boehler}, {Cabrit}, {Manara},
  {Menard}, {Mulders}, {Nisini}, {Fischer}, {Rigliaco}, {Banzatti}, {Avenhaus},
  \& {Gully-Santiago}}]{Long18}
{Long}, F., {Pinilla}, P., {Herczeg}, G.~J., {et~al.} 2018, \apj, 869, 17,
  \dodoi{10.3847/1538-4357/aae8e1}

\bibitem[{{Madhusudhan} {et~al.}(2012){Madhusudhan}, {Lee}, \&
  {Mousis}}]{Madhusudhan12}
{Madhusudhan}, N., {Lee}, K. K.~M., \& {Mousis}, O. 2012, \apjl, 759, L40,
  \dodoi{10.1088/2041-8205/759/2/L40}

\bibitem[{{Malik} {et~al.}(2019){Malik}, {Kitzmann}, {Mendon{\c{c}}a}, {Grimm},
  {Marleau}, {Linder}, {Tsai}, \& {Heng}}]{Malik19}
{Malik}, M., {Kitzmann}, D., {Mendon{\c{c}}a}, J.~M., {et~al.} 2019, \aj, 157,
  170, \dodoi{10.3847/1538-3881/ab1084}

\bibitem[{{Malik} {et~al.}(2017){Malik}, {Grosheintz}, {Mendon{\c{c}}a},
  {Grimm}, {Lavie}, {Kitzmann}, {Tsai}, {Burrows}, {Kreidberg}, {Bedell},
  {Bean}, {Stevenson}, \& {Heng}}]{Malik17}
{Malik}, M., {Grosheintz}, L., {Mendon{\c{c}}a}, J.~M., {et~al.} 2017, \aj,
  153, 56, \dodoi{10.3847/1538-3881/153/2/56}

\bibitem[{{Mbarek} \& {Kempton}(2016)}]{mbarek16}
{Mbarek}, R., \& {Kempton}, E. M.~R. 2016, \apj, 827, 121,
  \dodoi{10.3847/0004-637X/827/2/121}

\bibitem[{{Miller-Ricci Kempton} {et~al.}(2012){Miller-Ricci Kempton},
  {Zahnle}, \& {Fortney}}]{Kempton12}
{Miller-Ricci Kempton}, E., {Zahnle}, K., \& {Fortney}, J.~J. 2012, \apj, 745,
  3, \dodoi{10.1088/0004-637X/745/1/3}

\bibitem[{{Mishra} \& {Li}(2015)}]{Mishra15}
{Mishra}, A., \& {Li}, A. 2015, \apj, 809, 120,
  \dodoi{10.1088/0004-637X/809/2/120}

\bibitem[{{Morley} {et~al.}(2013){Morley}, {Fortney}, {Kempton}, {Marley},
  {Visscher}, \& {Zahnle}}]{morley13}
{Morley}, C.~V., {Fortney}, J.~J., {Kempton}, E. M.~R., {et~al.} 2013, \apj,
  775, 33, \dodoi{10.1088/0004-637X/775/1/33}

\bibitem[{{Mulders} {et~al.}(2021){Mulders}, {Dr{\k{a}}{\.z}kowska}, {van der
  Marel}, {Ciesla}, \& {Pascucci}}]{Mulders2021}
{Mulders}, G.~D., {Dr{\k{a}}{\.z}kowska}, J., {van der Marel}, N., {Ciesla},
  F.~J., \& {Pascucci}, I. 2021, \apjl, 920, L1,
  \dodoi{10.3847/2041-8213/ac2947}

\bibitem[{{{\"O}berg} {et~al.}(2011){{\"O}berg}, {Murray-Clay}, \&
  {Bergin}}]{Oberg11_C_O}
{{\"O}berg}, K.~I., {Murray-Clay}, R., \& {Bergin}, E.~A. 2011, \apjl, 743,
  L16, \dodoi{10.1088/2041-8205/743/1/L16}

\bibitem[{{Ohno} \& {Kawashima}(2020)}]{ohno20}
{Ohno}, K., \& {Kawashima}, Y. 2020, \apjl, 895, L47,
  \dodoi{10.3847/2041-8213/ab93d7}

\bibitem[{{Parmentier, Vivien} {et~al.}(2013){Parmentier, Vivien}, {Showman,
  Adam P.}, \& {Lian, Yuan}}]{Parmentier_2013}
{Parmentier, Vivien}, {Showman, Adam P.}, \& {Lian, Yuan}. 2013, A\&A, 558,
  A91, \dodoi{10.1051/0004-6361/201321132}

\bibitem[{Pavlov {et~al.}(2001)Pavlov, Brown, \& Kasting}]{Pavlov2001a}
Pavlov, A.~A., Brown, L.~L., \& Kasting, J.~F. 2001, Journal of Geophysical
  Research: Planets, 106, 23267, \dodoi{10.1029/2000JE001448}

\bibitem[{{Pearson} {et~al.}(2006){Pearson}, {Sephton}, {Franchi}, {Gibson}, \&
  {Gilmour}}]{Pearson06}
{Pearson}, V.~K., {Sephton}, M.~A., {Franchi}, I.~A., {Gibson}, J.~M., \&
  {Gilmour}, I. 2006, Meteoritics and Planetary Science, 41, 1899,
  \dodoi{10.1111/j.1945-5100.2006.tb00459.x}

\bibitem[{{Pollack} {et~al.}(1994){Pollack}, {Hollenbach}, {Beckwith},
  {Simonelli}, {Roush}, \& {Fong}}]{Pollack94}
{Pollack}, J.~B., {Hollenbach}, D., {Beckwith}, S., {et~al.} 1994, \apj, 421,
  615, \dodoi{10.1086/173677}

\bibitem[{{Qi} {et~al.}(2013){Qi}, {Oberg}, {Wilner}, {d'Alessio}, {Bergin},
  {Andrews}, {Blake}, {Hogerheijde}, \& {van Dishoeck}}]{Qi13_sci}
{Qi}, C., {Oberg}, K.~I., {Wilner}, D.~J., {et~al.} 2013, Science, 341, 630

\bibitem[{Rannou {et~al.}(1999)Rannou, McKay, Botet, \& Cabane}]{RANNOU1999}
Rannou, P., McKay, C., Botet, R., \& Cabane, M. 1999, Planetary and Space
  Science, 47, 385 , \dodoi{https://doi.org/10.1016/S0032-0633(99)00007-0}

\bibitem[{{Ros} \& {Johansen}(2013)}]{Ros13}
{Ros}, K., \& {Johansen}, A. 2013, \aap, 552, A137,
  \dodoi{10.1051/0004-6361/201220536}

\bibitem[{{Santerne} {et~al.}(2018){Santerne}, {Brugger}, {Armstrong},
  {Adibekyan}, {Lillo-Box}, {Gosselin}, {Aguichine}, {Almenara}, {Barrado},
  {Barros}, {Bayliss}, {Boisse}, {Bonomo}, {Bouchy}, {Brown}, {Deleuil},
  {Delgado Mena}, {Demangeon}, {D{\'\i}az}, {Doyle}, {Dumusque}, {Faedi},
  {Faria}, {Figueira}, {Foxell}, {Giles}, {H{\'e}brard}, {Hojjatpanah},
  {Hobson}, {Jackman}, {King}, {Kirk}, {Lam}, {Ligi}, {Lovis}, {Louden},
  {McCormac}, {Mousis}, {Neal}, {Osborn}, {Pepe}, {Pollacco}, {Santos},
  {Sousa}, {Udry}, \& {Vigan}}]{Santerne18}
{Santerne}, A., {Brugger}, B., {Armstrong}, D.~J., {et~al.} 2018, Nature
  Astronomy, 2, 393, \dodoi{10.1038/s41550-018-0420-5}

\bibitem[{{Seager} {et~al.}(2007){Seager}, {Kuchner}, {Hier-Majumder}, \&
  {Militzer}}]{Seager07}
{Seager}, S., {Kuchner}, M., {Hier-Majumder}, C.~A., \& {Militzer}, B. 2007,
  \apj, 669, 1279, \dodoi{10.1086/521346}

\bibitem[{{Stevenson} {et~al.}(2010){Stevenson}, {Harrington}, {Nymeyer},
  {Madhusudhan}, {Seager}, {Bowman}, {Hardy}, {Deming}, {Rauscher}, \&
  {Lust}}]{stevenson10}
{Stevenson}, K.~B., {Harrington}, J., {Nymeyer}, S., {et~al.} 2010, \nat, 464,
  1161, \dodoi{10.1038/nature09013}

\bibitem[{{St{\"o}kl} {et~al.}(2015){St{\"o}kl}, {Dorfi}, \&
  {Lammer}}]{Stokl15}
{St{\"o}kl}, A., {Dorfi}, E., \& {Lammer}, H. 2015, \aap, 576, A87,
  \dodoi{10.1051/0004-6361/201423638}

\bibitem[{{Swain} {et~al.}(2008){Swain}, {Vasisht}, \& {Tinetti}}]{swain08}
{Swain}, M.~R., {Vasisht}, G., \& {Tinetti}, G. 2008, \nat, 452, 329,
  \dodoi{10.1038/nature06823}

\bibitem[{{Tabone} {et~al.}(2023){Tabone}, {Bettoni}, {van Dishoeck},
  {Arabhavi}, {Grant}, {Gasman}, {Henning}, {Kamp}, {G{\"u}del}, {Lagage},
  {Ray}, {Vandenbussche}, {Abergel}, {Absil}, {Argyriou}, {Barrado},
  {Boccaletti}, {Bouwman}, {Garatti}, {Geers}, {Glauser}, {Justannont},
  {Lahuis}, {Mueller}, {Nehm{\'e}}, {Olofsson}, {Pantin}, {Scheithauer},
  {Waelkens}, {Waters}, {Black}, {Christiaens}, {Guadarrama},
  {Morales-Calder{\'o}n}, {Jang}, {Kanwar}, {Pawellek}, {Perotti}, {Perrin},
  {Rodgers-Lee}, {Samland}, {Schreiber}, {Schwarz}, {Colina}, {{\"O}stlin}, \&
  {Wright}}]{Tabone23}
{Tabone}, B., {Bettoni}, G., {van Dishoeck}, E.~F., {et~al.} 2023, Nature
  Astronomy, submitted, \dodoi{10.48550/arXiv.2304.05954}

\bibitem[{{Teal} {et~al.}(2022){Teal}, {Kempton}, {Bastelberger}, {Youngblood},
  \& {Arney}}]{Teal22}
{Teal}, D.~J., {Kempton}, E. M.~R., {Bastelberger}, S., {Youngblood}, A., \&
  {Arney}, G. 2022, \apj, 927, 90, \dodoi{10.3847/1538-4357/ac4d99}

\bibitem[{Tsai {et~al.}(2021)Tsai, Malik, Kitzmann, Lyons, Fateev, Lee, \&
  Heng}]{Tsai_2021}
Tsai, S.-M., Malik, M., Kitzmann, D., {et~al.} 2021, The Astrophysical Journal,
  923, 264, \dodoi{10.3847/1538-4357/ac29bc}

\bibitem[{{Unterborn} {et~al.}(2014){Unterborn}, {Kabbes}, {Pigott}, {Reaman},
  \& {Panero}}]{Unterborn14}
{Unterborn}, C.~T., {Kabbes}, J.~E., {Pigott}, J.~S., {Reaman}, D.~M., \&
  {Panero}, W.~R. 2014, \apj, 793, 124, \dodoi{10.1088/0004-637X/793/2/124}

\bibitem[{Wilson \& Atreya(2003)}]{WILSON_2003}
Wilson, E., \& Atreya, S. 2003, Planetary and Space Science, 51, 1017 ,
  \dodoi{https://doi.org/10.1016/j.pss.2003.06.003}

\bibitem[{{Wolf} \& {Toon}(2010)}]{Wolf_2010}
{Wolf}, E.~T., \& {Toon}, O.~B. 2010, Science, 328, 1266,
  \dodoi{10.1126/science.1183260}

\bibitem[{{Zahnle} {et~al.}(2016){Zahnle}, {Marley}, {Morley}, \&
  {Moses}}]{Zahnle16}
{Zahnle}, K., {Marley}, M.~S., {Morley}, C.~V., \& {Moses}, J.~I. 2016, \apj,
  824, 137, \dodoi{10.3847/0004-637X/824/2/137}

\bibitem[{{Zeng} {et~al.}(2019){Zeng}, {Jacobsen}, {Sasselov}, {Petaev},
  {Vanderburg}, {Lopez-Morales}, {Perez-Mercader}, {Mattsson}, {Li}, {Heising},
  {Bonomo}, {Damasso}, {Berger}, {Cao}, {Levi}, \& {Wordsworth}}]{Zeng19}
{Zeng}, L., {Jacobsen}, S.~B., {Sasselov}, D.~D., {et~al.} 2019, Proceedings of
  the National Academy of Science, 116, 9723, \dodoi{10.1073/pnas.1812905116}

\end{thebibliography}
\bibliographystyle{aasjournal}



\end{document}